\documentclass[aps,pre,twocolumn,showpacs]{revtex4-1}
\usepackage{amsmath,amsfonts,bm, color,amssymb}
\usepackage{graphicx}

\newcommand{\kT}{{\mathrm  k_BT}}
\newcommand{\out}{{\mathrm{ex}}}
\newcommand{\ins}{{\mathrm{in}}}
\newcommand{\im}{{\mathrm{i}}}
\newcommand{\refeq}[1]{Eq.(\ref{#1})}
\newcommand{\col}[1]{{\color{black}{#1}}}
\newcommand{\reffig}[1]{Figure (S\ref{#1})}

\begin{document}

\title{Fluctuation spectroscopy of giant unilamellar vesicles using confocal and phase contrast microscopy} 

\author{Hammad A. Faizi$^{a}$, Cody J. Reeves$^{c}$, Vasil N. Georgiev$^{b}$, Petia M. Vlahovska$^{\ast, c}$ and Rumiana Dimova$^{\ast, b}$}
\affiliation{
$^{a}$~Department of Mechanical Engineering, Northwestern University, Evanston, IL 60208, USA\\
$^{b}$~Department of Theory and Biosystems, Max Planck Institute of Colloids and Interfaces, Science Park Golm, 14424 Potsdam, Germany.\\
$^{c}$~ Department  of Engineering Sciences and Applied Mathematics, Northwestern University, Evanston, IL 60208, USA. \\
\textit{$*$ E-mails for correspondence Rumiana.Dimova@mpikg.mpg.de, Petia.Vlahovska@northwestern.edu}}

\date{\today}

\begin{abstract}

{A widely used method  to measure the bending rigidity of bilayer membranes is  fluctuation spectroscopy, which analyses the thermally-driven membrane undulations of giant unilamellar vesicles recorded with either phase-contrast or confocal microscopy.  Here, we analyze the fluctuations of the same vesicle using both techniques  and  obtain consistent values for the bending modulus. We discuss the factors that may lead to discrepancies. }

\end{abstract}

\maketitle


Bending rigidity of cellular membranes plays a key role in membrane remodeling. Knowledge of its  value is needed to quantify processes that involve curvature changes such as budding (as in endo- and exocytosis), tubulation and fusion. Various experimental methods have been devised to measure bending rigidity\cite{DIMOVA2014225}, e.g. micropipette aspiration \cite{Evans-Rawicz:1990, RAWICZ2000328}, electrodeformation \cite{Kummrow-Helfrich:1991,Vlahovska-Dimova:2009,gracia.2010,Salipante-Vlahovska:2012}, optical tweezers \cite{TIAN20091636, Sorre:2009}, and scattering based techniques \cite{KUCERKA2005, PAN20122135}.
 One of the most popular methods is fluctuation spectroscopy, pioneered by \textit{Brochard and Lenon}\cite{brochard.1975}, due to its ease of implementation\cite{mitov:1989,Pecreaux:2004,gracia.2010,genova.2013}.  In essence, a time series of vesicle contours in the focal plane (the equator of the quasi-spherical vesicle) is recorded. The quasi-circular  contour is decomposed in Fourier modes. The  fluctuating amplitudes have variance dependent on the membrane bending rigidity and  tension.  Imaging is most commonly done by  phase contrast microscopy \cite{Pecreaux:2004,gracia.2010,DIMOVA2014225, BASSEREAU201447, faizi2019, Bouvrais18442, elani.2015, Almendro-Vedia11291, Meleard:2011}  but other methods such as  confocal \cite{DRABIK2016244, dahl.2016, rautu.2017} and light sheet microscopy \cite{loftus2013} have also been employed. The increased variety of imaging methods raises the question whether they all yield the same results.

Recently, \textit{Rautu et al.}\cite{rautu.2017} pointed out that in phase contrast  imaging, projections of out-of-focus fluctuations may contribute to the contour statistics leading to systematic overestimation of the bending rigidity value when compared to other methods such as micropipette aspiration and X-ray scattering.  However, comparing bending rigidity numbers obtained by different techniques is only meaningful if the same system is probed. It is known that many factors such as sugars (and gravity), salt, buffers, solution asymmetry, concentration of fluorescent lipids, preparation method or type of bilayer configuration (stacked or free-floating), influence the measured mechanical properties of bilayer membranes \cite{DIMOVA2014225,vitkova2006, faizi2019, karimi2018,Henriksen2002, BOUVRAIS20101333, Bouvrais2014}; {see Table 1 in the Supporting Information (SI) for a list of reported bending rigidity of DOPC membranes. For example, even measurements with the same method can give a wide range of values, e.g., the bending rigidity of a DOPC bilayer measured with flickering spectroscopy has been reported from 15 $\kT$\cite{dahl.2016} to 30 $\kT$ \cite{elani.2015}, where $\kT$\ is the thermal energy.

 In order to compare imaging with phase contrast and confocal microscopy, which was suggested in \textit{Rautu et al.}\cite{rautu.2017} as a better technique due to the precise control over the focal depth, we measure the bending rigidity of the same giant vesicle with both techniques. We highlight some important issues to be considered to ensure reliable measurements. We also show that results obtained with both methods are consistent.

\subsection*{Equilibrium fluctuations of a quasi--spherical vesicle}

 First, we summarize the theoretical basis of the fluctuations analysis  
 (details are provided in SI section {S6}). We also correct  published expressions for the  relaxation frequency and cross-spectral density of the shape fluctuations.
 
The contour in the equatorial plane of a quasi-spherical vesicle is decomposed in Fourier modes, 
$r(\phi,t)=R_0\left(1+\sum_{q=-q_{\max}}^{q_{\max}}u_q(t) \exp(\im q \phi)\right)$, where $R_0=(3V/4{\pi})^{1/3}$ is the radius of an equivalent sphere with the volume $V$ of the GUV and $q$ is the mode number. 
 In practice, $q_{\max}$ is the maximum number of experimentally resolved modes. The statistical analysis of the fluctuating  amplitudes $u_q$ yields the values of  membrane bending rigidity $\kappa$ and the tension $\sigma$ since
$\langle \left|u_{q}\right |^2\rangle\sim {\kT}/{\kappa \left(q^3+\bar \sigma q\right)}$, where $\bar\sigma=\sigma R_0^2/\kappa$.

More precisely, the statistics of the two-dimensional circular modes, $u_q$, is derived from the  three-dimensional shape modes, $f_{lm}$, which describe the nearly-spherical shape in terms of spherical harmonics  \cite{mitov:1989, Seifert:1999},
$      R(\theta,\phi,t) = R_0\left(1+ \sum_{l=0}^{l_{\max}} \sum_{m=-l}^{l} f_{lm} (t) \ \mathcal{Y}_{lm} \big(\theta,\phi\big)\right)\,$.
Here, $l_{\max}$ is an upper cutoff, in the order of the ratio of the GUV radius and bilayer thickness.  The contour in the focal plane corresponds to the equator of the quasi-spherical vesicle, $\theta=\pi/2$, i.e., $r(\phi,t)=R(\frac{\pi}{2}, \phi,t)$, which leads to the following expression for the mean squared amplitudes 
 \begin{equation}
  \label{Hs1}
  \langle |u_q|^2\rangle=\frac{\kT}{\kappa}\sum_{l=q}^{l_{max}}{n_{lq} P^2_{lq}(0)} \left[(l-1)(l+2)\left(l(l+1)+\bar  \sigma\right)\right]^{-1}
 \end{equation}
where $n_{lq}=(2l+1)(l-q)!/4\pi (l+q)!$ and $P_{lq}$  are the associated Legendre polynomials.
 The short-wavelength 
 shape fluctuations are dominated by the bending rigidity, while the long wavelengths are controlled by tension; the crossover occurs around mode ~$q_c=\sqrt{\bar{\sigma}}$.

To validate our methodology, we have simulated the thermal  shape fluctuations of a GUV, see also SI section S6. We have generated a sequence of three-dimensional shapes (and their corresponding equatorial contours)  using the evolution equations  \cite{Seifert:1999,Miao:2002}
\begin{equation}
 \label{eqF}
 \begin{split}
\frac{d f_{lm}}{d t}=&-\tau_l^{-1} f_{lm}+\zeta_{lm}(t)\,,\\
\tau_l^{-1} =& \frac{\kappa}{\eta_\out R_0^3} \frac{(l-1)l(l+1)(l+2)\left(l(l+1)  + \bar\sigma \right)
}{4l^3 +6l^2 -1+\left(2l^3+3l^2-5\right)\left(\frac{\eta_\ins}{\eta_\out}-1\right)} \\
\end{split}
 \end{equation}
 where $\zeta_{lm}$ is the thermal noise driving the membrane undulations, 
 $\eta_\ins$ and $\eta_\out$ are the viscosity of the solution inside and outside the vesicle.  Note that the relaxation time given by Eq. 2 in  \textit{Rautu et al.}\cite{rautu.2017} has incorrect dependence on the viscosities of the enclosed and suspending solutions (this mistake is unlikely to affect the reported fluctuation spectra). The simulated contours were analyzed by our code and the extracted bending rigidity and tension were compared to the input values to confirm accuracy of the contour detection, Fourier decomposition and data fitting algorithms. The time evolution of the modes also enables us to access information provided by the time correlations
$ \langle u_q(0)u^*_q(t)\rangle= \langle |u_q|^2\rangle \exp(-t/\tau_q)\,.$
 If $q\gg1$, the correlation time tends to that of a planar membrane $\tau_q^{-1}=\kappa \left(q^3+\bar \sigma q\right)/2R_0^3(\eta_\out+\eta_\ins)$.
If the cross-spectral density $\langle |u_q(0)||u_q(t)|\rangle$ is utilized, the correct time dependence in the exponential includes a factor of 2 and Eq. 3 in \textit{Zhou et al.} \cite{Losert:2011} needs to be corrected (see SI section S6, Eqn.  39).

\begin{figure}[h]
\centering
 \includegraphics[width=1.2\linewidth]{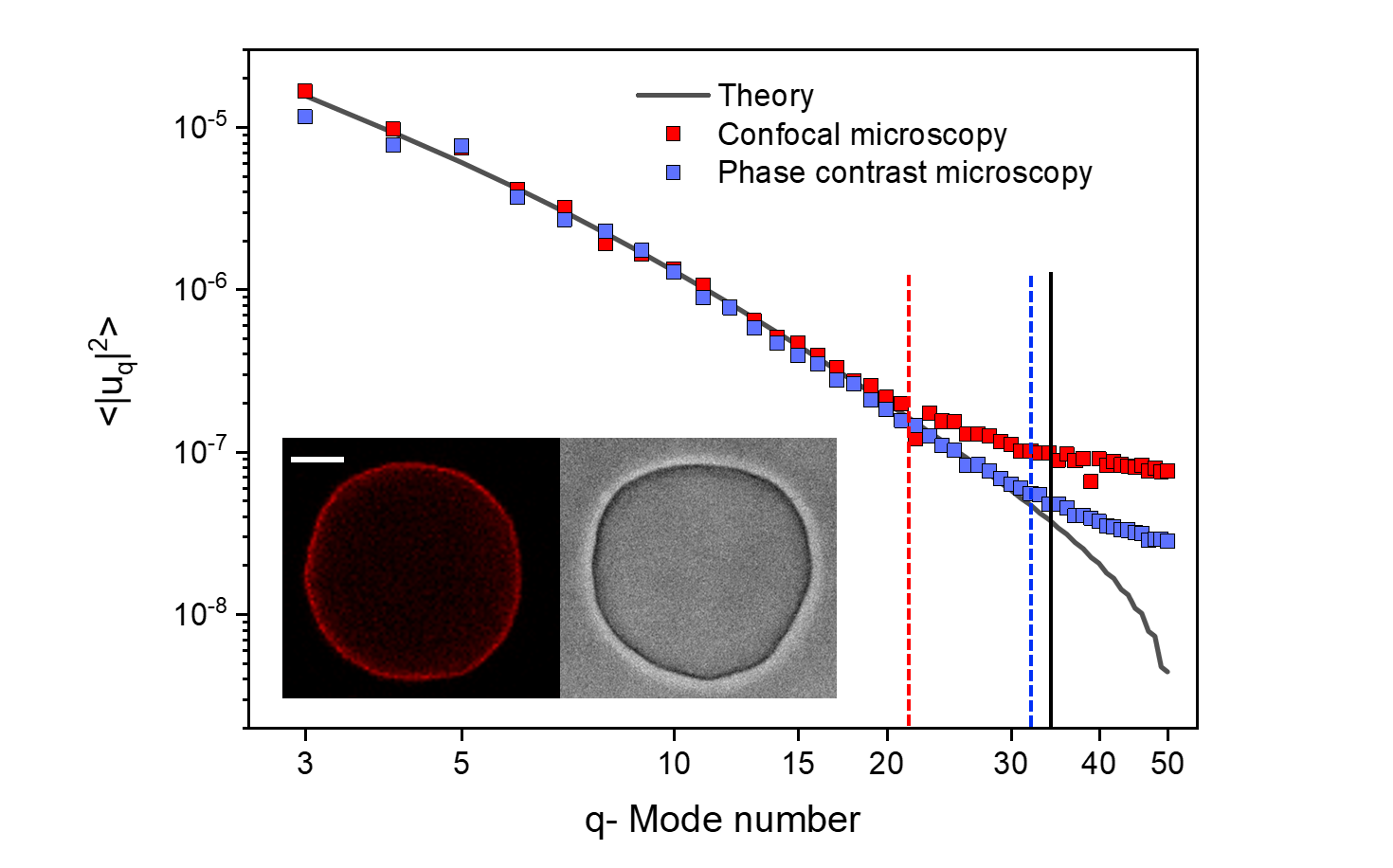}
  \caption{
  Fluctuation spectrum of the same DOPC vesicle (shown in the inset) obtained with confocal and phase contrast microscopy with a 40x/NA 0.6 objective, pinhole size of 1 AU and polarization correction, see also supplementary Movies S1 and S2. The dye concentration is 0.2 mol$\%$. Scale bar is 15 $\mu$m. The vertical lines denote the cutoff resolution for the modes: optical resolution (solid line), phase contrast (blue dashed) and confocal (red dashed line). The theory is fitted to phase contrast data up to the mode marked by the dashed blue line.
The crossover mode $q_C=\sqrt{\bar\sigma}$ is 7.
  }
  \centering
  \label{figure1}
\end{figure}

\subsection*{Bending rigidity obtained from confocal and phase-contrast microcopy: effect of resolution and vesicle size}

Giant unilamellar vesicles (GUV) were electroformed  from DOPC (99.8 mol\% dioleoylphosphatidylcholine and 0.2 mol\% Texas-Red 1,2-hexadecanoyl-wn-glycero-3-phosphoethanolamine, TR-DHPE) in 20 mM sucrose and subsequently diluted in 22 mM glucose, see SI section S2 for details.  Low sugar concentration was used in order to minimize the effects of  gravity\cite{Henriksen2002} and effect of sugars\cite{vitkova2006}, but still allow the vesicles to settle to the chamber bottom for easier recording. Low dye content minimizes effects of fluorophores\cite{BOUVRAIS20101333}.

Figure \ref{figure1} shows a typical fluctuations spectrum, given by \refeq{Hs1}, fitted to the experimental data for the same vesicle imaged with confocal and phase contrast microscopy using a 40x objective with 0.6 numerical aperture (NA), pinhole size of 1 Airy unit (AU) and polarization correction (see below and SI section S3). The contour was detected with sub-pixel resolution \cite{gracia.2010}. The experimental data was fitted with \refeq{Hs1} with Levenberg-Marquardt algorithm and yielded  bending rigidity $\kappa= 23.9\pm1.6\, \kT$ and tension $\sigma =5.1\pm1.4\times 10^{-9}$ N/m and $\kappa=22.3\pm 2.1\,\kT$ and  $\sigma=3.1\pm1.2\times10^{-9}$ N/m  from the confocal and phase contrast microscopy data, respectively. The average and error in individual GUV was determined by performing fluctuation spectroscopy 2-3 times. By imaging a population of 18 vesicles with both methods, the bending rigidity obtained are 22.5$\pm$2.0 $\kT$ with confocal and 23.3$\pm$1.6 $\kT$ phase contrast microscopy; each vesicle was analyzed with both imaging techniques as in Fig. \ref{figure1} and then the results were averaged over the population.  Figure \ref{figure2} shows the box and whisker plot for more detailed statistics. Based on F statistics and ANOVA (analysis of variance) test, the $p$-value obtained is $p=0.48$ for null hypothesis testing. We also performed the paired-sample t-test and obtained $p=0.43$. These $p$-values indicate no significant difference between the two imaging techniques.

\begin{figure}[h]
\centering
\includegraphics[width=1.15\linewidth]{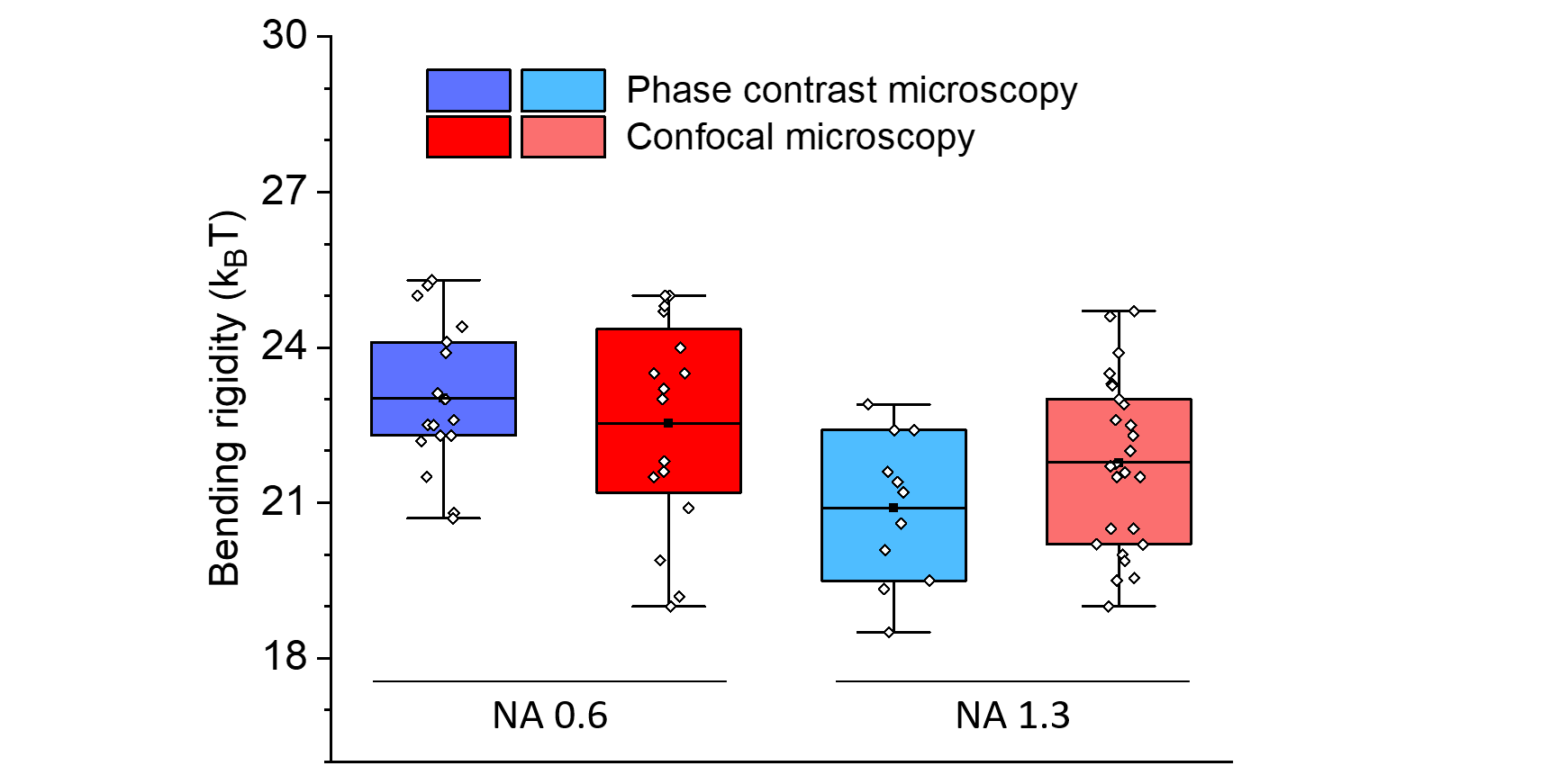}
  \caption{
Imaging with phase contrast and confocal microscopy for objectives of the same numerical aperture (NA) give consistent results. Box and whisker plot comparison for a DOPC vesicle population where each vesicle was analyzed with phase contrast and confocal imaging with 40x objectives with NA 0.6 and NA 1.3. Pinhole size is 1 AU with polarization correction. The dye concentration is 0.2 mol$\%$.}
  \label{figure2}
\end{figure}

Since only modes with wavenumber $q>\sqrt{\sigma R_0^2/\kappa}$ are sensitive to the bending rigidity, 
it is desirable to have more resolved modes, i.e., modes with amplitude and wavelength greater than the optical resolution limit $\approx 250$ nm \cite{LIPOWSKY1995521}.   The average mean fluctuation amplitude scales with the size of the vesicle $u_q\sim R_0\sqrt{\kT/\kappa}$, hence  larger GUVs admit more spatially resolvable fluctuation modes as shown in Fig. \ref{figure3}. However, even for the same vesicle we find that the number of  resolved modes is higher for phase contrast than for confocal imaging. Indeed, Fig. \ref{figure1} shows that the noise level is higher for confocal microscopy, and on average phase contrast imaging resolves 8-10 modes more than confocal imaging does. The poorer mode resolution with confocal microscopy is likely due to poor contour recognition.  The reasons for this are discussed in the next section.

\begin{figure}[h]
\centering
 \includegraphics[width=1.2\linewidth]{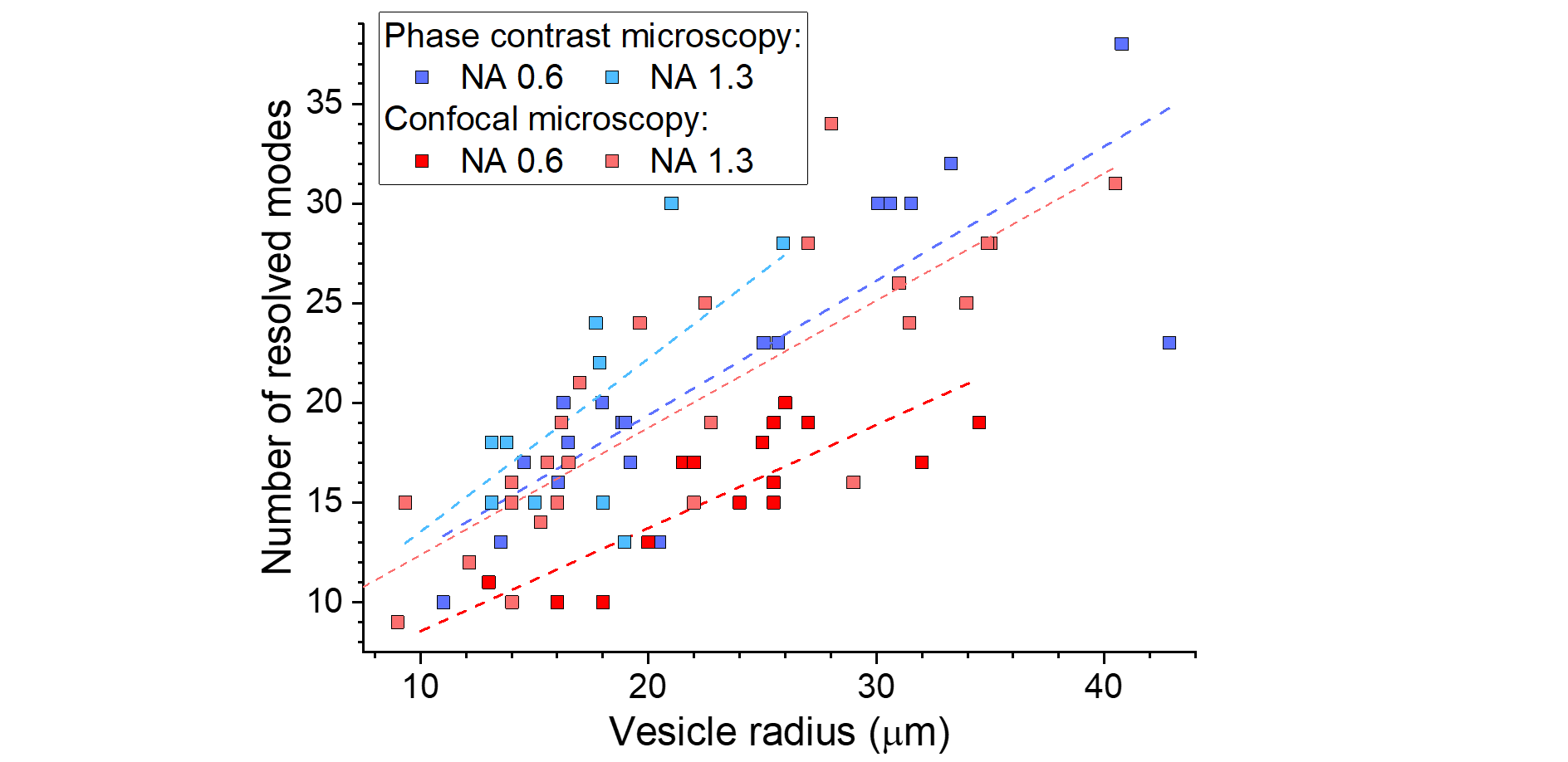}
  \caption{
Larger vesicles allow resolving more fluctuation modes thus yielding more reliable determination of the bending rigidity. Data are collected on DOPC vesicles with different sizes. The dye concentration is 0.2 mol$\%$. Regression lines are added to guide the eyes. Imaging was done with 40x objectives with different numerical aperture (NA), pinhole size of 1 AU and polarization correction. 
  }
  \label{figure3}
\end{figure}
We found that the vesicle population needs to have broad  range of radii  to avoid a size bias in the bending rigidity values we discovered for confocal microscopy with low resolution optics, see also SI section S4. In the case of 40x/NA 0.6 (air) objective, the mean Pearson correlation and standard deviation coefficient is 0.65$\pm0.21$ (see SI section S5 for the histograms generated with bootstrapping resampling technique). Analysis of a population of similar sized vesicles with radii around 10 $\mu$m 
underestimates $\kappa$ by roughly 6 $\kT$. The bias originates from out-of-plane fluorescence which worsens the contour detection. This issue is investigated in the next section and SI section 6.
The size dependence is insignificant for phase contrast microscopy with a mean correlation coefficient of 0.28$\pm$0.18 with 40x/NA 0.6 (air). Analyzing the same vesicle population with 40x/NA 1.3 objective in phase contrast and confocal imaging yields 21.0$\pm$2.0 $\kT$ and 21.7$\pm$2.0 $\kT$ respectively. Higher numerical aperture in phase contrast  leads to negligible correlation coefficient of -0.07$\pm$0.34 between bending rigidity and vesicle size and decrease in the correlation coefficient to 0.43$\pm$0.14 for confocal imaging with 40x/NA 1.3 objective.


\subsection*{Out-of-focus fluorescence affects contour detection quality in confocal microscopy }

The vesicle contour is detected from radial intensity line profiles, see SI section S2. In confocal cross sections, weak fluorescence from the vesicle membrane located above and below the focal plane may result in signal projected in the interior of the vesicle image which is higher compared to the surrounding background. The resulting asymmetry in the intensity line profile (Fig.\ref{figure4}a) leads to an artificial contour displacement, i.e., poor contour detection (note that such an asymmetry is absent in images acquired with phase contrast microscopy of vesicles with similar refractive indices of the internal and external solutions). This asymmetry creates a systematic error shifting the vesicle contour by 0.53 $\mu$m. The error is larger than the pixel resolution of the system, 0.252  $\mu$m, hence the higher modes are averaged out. Smaller vesicles or larger pinholes lead to higher signal inside the vesicle (see inset in Fig. \ref{figure4}b) corresponding to greater asymmetry which increases the error from contour fitting and introduces dependence of the bending rigidity on vesicle size. For imaging with higher numerical aperture objectives (e.g. NA 1.3), the asymmetry in the intensity line profiles is suppressed and contour detection is correct. Note that phase contrast images do not suffer from the asymmetry-induced error irrespective of the objective NA.  

\begin{figure}[h]
\centering
\includegraphics[width=\linewidth]{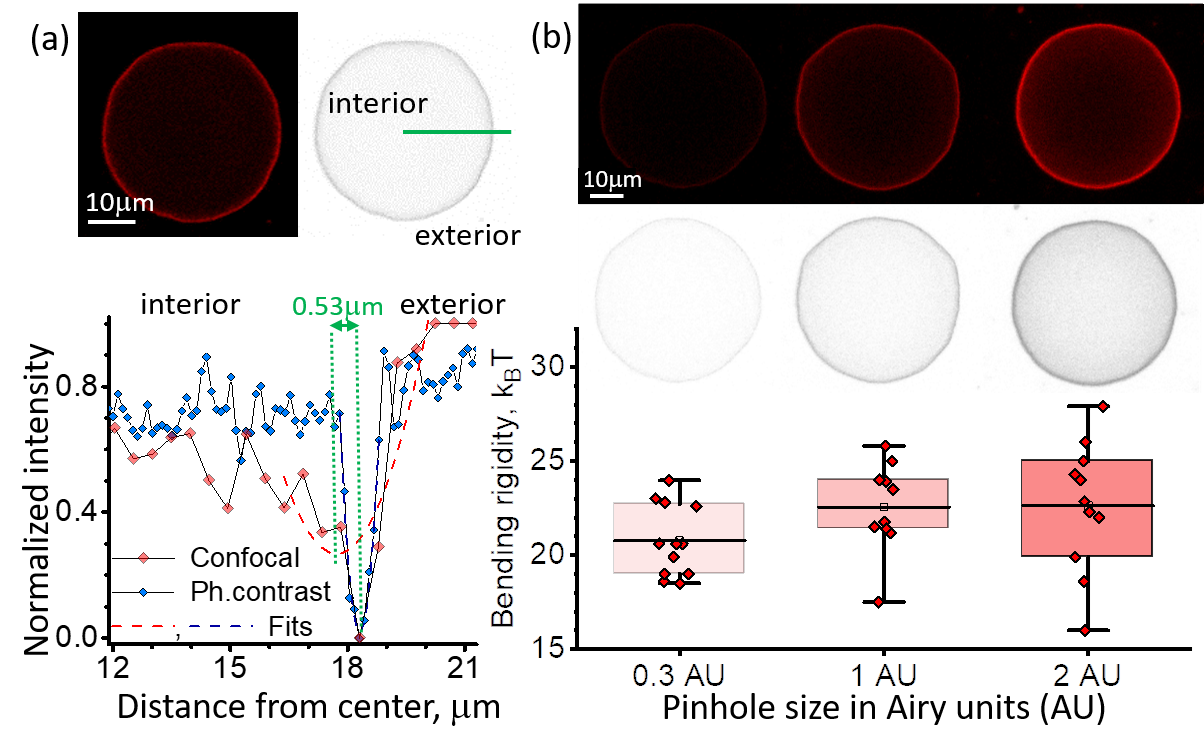}
  \caption{ Out-of-focus fluorescence in confocal images can result in erroneous contour detection and increased error in bending rigidity. (a) Intensity line profiles (gray-value) across the vesicle membrane (DOPC, with 0.2 mol$\%$ dye) are symmetric for phase-contrast images (blue) but asymmetric for confocal images (red, 1 AU). The asymmetry in the confocal line profile leads to incorrect detection of the contour position defined by the parabolic fit minimum, here, shifted inwards by 0.53 $\mu$m. (b)  Vesicle images (and their inverted gray-value analogs) acquired with different pinhole size show increased fluorescence inside the vesicle which results in larger error in the bending rigidity. Box and whisker plot of the bending rigidity of the same DOPC vesicles imaged with confocal microscopy at three different pinhole sizes for 40x/NA 0.6 objective and polarization correction.
 }
  \label{figure4}
\end{figure}

We investigated the impact of out-of-focus fluorescence on the  fluctuations statistics by varying the pinhole size for confocal imaging on the same vesicle. The standard pinhole size in confocal microscopy is defaulted to 1 Airy unit (AU) (full width at half maximum FWHM=1.6 $\mu$m) for 40x/NA 0.6 objective. We analyzed the same vesicles with different optical sectioning at 0.3 AU (FWHM=0.9 $\mu$m), and 2 AU  (FWHM=2.9 $\mu$m). The mean bending rigidity did not show significant differences based on ANOVA testing, post hoc Dunnett test and paired-sample t-test (p=0.87), however the error increases with the pinhole size. The sensitivity to the vesicle size also becomes more pronounced with higher pinhole size. At the  largest pinhole size (2.0 AU) the Pearson correlation coefficient is 0.60$\pm$0.22, while for 0.3 AU it becomes negligible,  -0.14$\pm$0.30. 

\subsection*{Dye related artifacts: vesicle tubulation and polarization}

Confocal imaging relies on fluorophores added to the membrane, and some studies  have used up to 10 mol\% dye \cite{dahl.2016,DRABIK2016244}. 
To probe the effect of fluorophore on $\kappa$,  we changed the dye concentration from 0.2 mol\% to 2 mol\% TR-DHPE. The bending rigidity of this population of vesicles showed non significant difference with $\kappa$=20.09$\pm$2.49 $\kT$ with one ANOVA testing. However, it was observed that over 2-3 min of recording, around 50 \% of the vesicles  developed inward structures such as buds or visible tubes as shown in Fig. \ref{figure5}. Vesicles with such defects displayed significantly higher  bending rigidity, $25.01\pm2.11$  $\kT$.

\begin{figure}[h]
\centering
  \includegraphics[width=\linewidth]{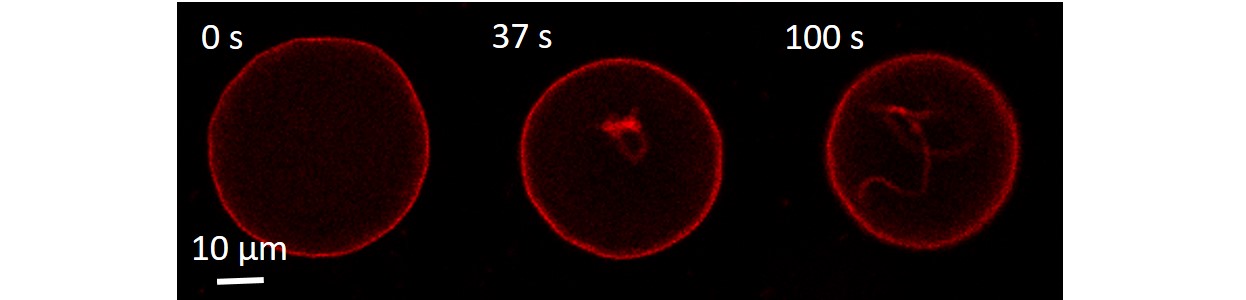}
  \caption{Time lapse of a DOPC vesicle with 2 mol\% TR-DHPE developing inward nanotubes as a result of long exposure to laser during confocal imaging. The second and third cross sections are non-equatorial to better show the formed nanotubes. 
}
  \label{figure5}
\end{figure}

TR-DHPE belongs to a family of polarity-sensitive fluorescent probes.
As a result, the signal intensities are different at the pole and equator of the vesicle (see SI section S3). This may  lead to errors in the contour detection in these regions. 
The polarization effect was corrected by using circular rotation plates to have even intensities across the equatorial vesicle plane. The analysis of the same vesicle with and without the polarization correction showed a 3 $\kT$ lower bending rigidity without  any correction with 40x/NA 0.6 (air) objective. This softening effect became insignificant with 40x/NA 1.3 (oil) objective (SI section S3). This is likely due to  loss of signal at low intensity regions where the higher mode fluctuations intensities are averaged out with background noise due to out-of-focus  fluorescence.

\subsection*{Effect of nearby vesicles on fluctuation spectra}
The equilibrium shape  fluctuations of an isolated GUV are driven by Gaussian thermal noise. Defects such as buds, nanotubes, invaginations or docked LUVs modify the vesicle fluctuations \cite{gracia.2010} and their effect can be detected in the statistics at each point on the vesicle contour profile using the ensemble-averaged probability density function (PDF) as shown in Fig. \ref{figure6}a 
In addition to defects attached  the membrane, we also found that hydrodynamic flows and/or fluorescence signal from nearby vesicles can affect vesicle fluctuations.

We characterized the Gaussianity of  the fluctuations using the fourth PDF moment, Kurtosis, $K$. For a Gaussian distribution, $K=3$. In Fig. \ref{figure6} we demonstrate how thermal fluctuations may be modified \textcolor{black} {(see supplementary Movie S3)}. As shown, the majority of contours are characterized by a normal distribution. However near other flickering structures, the fluctuation map density is modified. The non-Gaussian enhanced fluctuations are observed with leptokurtic nature ($K>3$). This observation serves as a caution to filter out vesicles with sub-optical structures affecting the fluctuations.

\begin{figure}[h]
\centering
 \includegraphics[width=\linewidth]{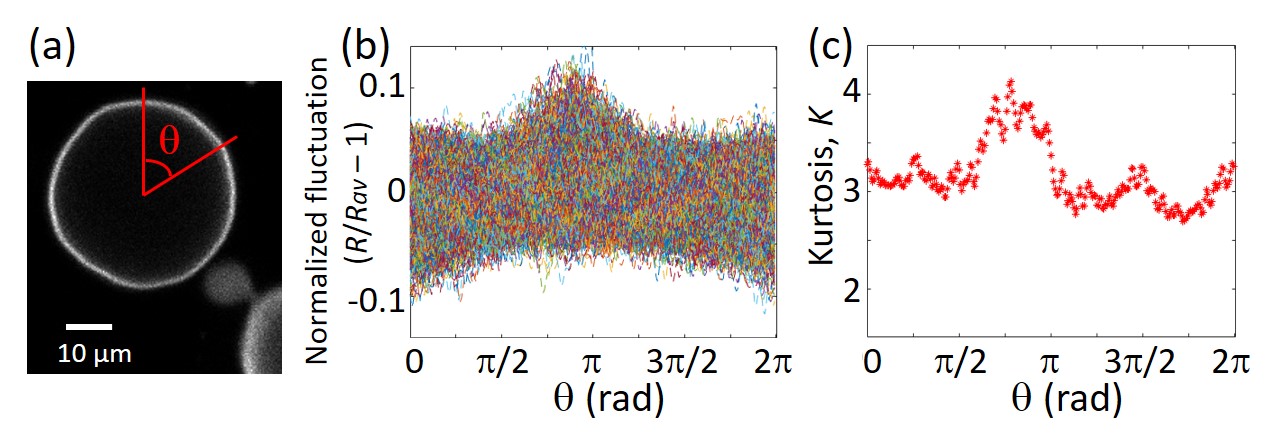}
  \caption{Nearby structures affect the fluctuation spectrum. (a) A flickering vesicle in close proximity to another vesicle bud. (b) Fluctuation density map of the vesicle in (a): the fluctuations are modified by hydrodynamic interactions of the other flickering vesicle bud. (c) Kurtosis $K>3$ indicates the vesicle fluctuations have amplified meaning local softening of the membrane.}
  \label{figure6}
\end{figure}

\subsection*{Conclusions}

We compare  the bending rigidity of bilayer membranes determined from flickering spectroscopy of GUVs imaged with confocal and phase contrast microscopy. Examining the same vesicle with both imaging techniques shows no significant differences in the bending modulus obtained from the two methods, in contrast to the  overestimation  reported by Rautu et al \cite{rautu.2017} when phase contrast microscopy is used. 
 Our analysis indicates that membrane defects  such as buds and tubes induced by  long laser exposure  in confocal microscopy can significantly stiffen the membrane.  Furthermore, we find that errors in contour detection  that could impact data interpretation can arise from fluorescence signal "pollution" and dye polarization.  
 The bending rigidity we obtain  ($\sim\,22\kT $ for DOPC) is in line with the values obtained with other techniques such as micropipette aspiration, X -ray scattering, electrodeformation and neutron spin echo \cite{RAWICZ2000328,TIAN20091636,KUCERKA2005}. A scatter of approximately 2 $\kT$ is typical in the experiments and should be taken into account when comparing data from different groups and methods.
 Exploring the effect of various parameters, we find that optimal imaging conditions for bending rigidity measurements from confocal imaging include high magnification objective, high numerical aperture, circular polarization correction, minimum dye concentration, small pinhole size, and broad vesicle size distribution.  
 
In conclusion, we demonstrate that phase contrast and confocal microscopy produce the same results if precautions are taken to minimize effects of the dye and improve contour detection. Our study suggests that the many published results obtained by phase contrast microscopy are likely to be unaffected by the projections  of out-of-focus fluctuations onto the imaging plane  in contrast to the claim by  Rautu et al \cite{rautu.2017}.  Since dye related artifacts such as laser-induced defects can compromise the data, it is advantageous to use phase contrast imaging  as it does not require dyes.

\section*{Acknowledgements}
This research was funded in part by NSF-CMMI awards 1748049 and 1740011. PV  acknowledges support from the Alexander von Humboldt Foundation  and HAF  acknowledges financial support from Prof. Reinhard Lipowsky for visits to the Max Planck Institute of Colloids and Interfaces. We thank Paul Salipante for proofreading the manuscript.  



\appendix

\begin{widetext}
\section*{S1. Bending rigidity value of DOPC bilayers}

The bending rigidity values of  bilayer membranes made of the same lipid can vary across studies due to different conditions, e.g., sugars, salt, buffers, dye concentration, as well as the preparation method \cite{DimovaACIS:2014}. Table \ref{diffcondmethod} illustrates the wide range of reported values of the bending rigidity values DOPC bilayers. {Refer to Table \ref{diffcondmethodhere} for the bending rigidity values obtained in this study for different microscopy setting}.
\begin{table*}[h]
\small
  \caption{Different bending rigidity values for DOPC under different conditions and methods.\textcolor{black}{PC, C and EP refer to phase contrast, confocal and epi-flourescent microscopies used respectively in Fluctuation spectroscopy. }}
  \label{diffcondmethod}
  \begin{tabular*}{\textwidth}{@{\extracolsep{\fill}}llllll}
    \hline
    Method & Rigidity ($\kT$) & Dye conc. ($\%mol$) & Buffer, Sugar (inside/outside) & Salt&  Preparation\\
    \hline
    Fluctuation Spec. [EP,C] & 14.9$\pm$0.4 \cite{dahl.2016} & 15.8 NBD PC & 100 mM Sucrose/100 mM Sucrose & N/A & Electroformation\\
    Fluctuation Spec. [PC] & 26.4$\pm$2.4 \cite{Gracia:2010} & 0 or 0.1 diIC18 & 10 mM Sucrose/10 mM Glucose & 0.1 mM NaCl & Electroformation\\
    Fluctuation Spec. [PC] & 26.8$\pm$2.4 \cite{elani.2015} & 1.0 Liss Rhod PE & 450 mM Sucrose/500 mM Glucose & N/A & Electroformation\\
    Fluctuation Spec. [PC] & 29.8$\pm$2.4 \cite{elani.2015} & 1.0 Liss Rhod PE & 450 mM Sucrose/500 mM Glucose & N/A & Phase Transfer\\
    Fluctuation Spec. [EP] & 22.3$\pm$0.5 \cite{kumar2020} & 0.12 Liss Rhod PE & 100 mM Sucrose/200 mM Sucrose & N/A & Electroformation\\
    Fluctuation Spec. [PC] & 27.3$\pm$3.2 \cite{brown2011} & N/A & 100 mM Sucrose/100 mM Sucrose & 2 mM NaN$_3$ & Electroformation\\
    Fluctuation Spec. [PC] & 22.7$\pm$2 \cite{tyler2019} & N/A & 100 mM Sucrose/125 mM Glucose & N/A & Electroformation\\
    Fluctuation Spec. [PC] & 21.46$\pm4$ \cite{Purushothaman2015} & N/A & 100 mM Sucrose/125 mM Glucose & N/A & Electroformation\\
    Fluctuation Spec. [PC] & 19$\pm$1 \cite{Shchelokovskyy_2011} & N/A & 10 mM Sucrose/10 mM Glucose & N/A & Electroformation\\
    Fluctuation Spec. [C] & 19$\pm$1 \cite{Rautu:2017} & 0.8 TR DHPE & 197 mM Sucrose/200 mM Glucose & N/A & Electroformation\\
    Time Correlations & 22.1 \cite{Losert:2011} & N/A & 300 mM Sucrose/307 mM Glucose & N/A & Electroformation\\
    Micropipette Aspiration & 20.7$\pm$2 \cite{Rawicz:2000} & N/A & 100 mM Sucrose/100 mM Glucose & N/A & Thin Film Hyd.\\
    Micropipette Aspiration & 22.8$\pm$2.2 \cite{Shchelokovskyy_2011} & N/A & 8 mM Sucrose/8 mM Glucose & N/A & Electroformation\\
    X- Ray Scattering & 20$\pm$2 \cite{Pan2008} & N/A & Water/Water & N/A & Extrusion\\
    X- Ray Scattering & 20.2$\pm$1.4 \cite{Jablin2014} & N/A & Deionized water & N/A & Bilayer stack\\
    Electrodeformation & 21.9$\pm$2 \cite{Gracia:2010} & 0.1 diIC18 & 10 mM Sucrose/10 mM Glucose & 0.1 mM NaCl & Electroformation\\
    Tether pulling & 20$\pm$2 \cite{Sorre:2009} & 0.3 TR DHPE & 300 mM Sucrose/80 mM Glucose & 100 mM NaCl & Electroformation\\
    Neutron Spin Echo & 20$\pm$1 \cite{Gupta2018} & N/A & D$_2$O/D$_2$O & 0 mM & Extrusion\\
    Neutron Spin Echo & 20$\pm$2 \cite{mel2020influence} & N/A & D$_2$O/D$_2$O & 0 mM & Extrusion\\
    Neutron Spin Echo & 30$\pm$4 \cite{mel2020influence} & N/A & D$_2$O/D$_2$O & 150 mM & Extrusion\\
    Neutron Spin Echo & 40$\pm$5 \cite{mel2020influence} & N/A & D$_2$O/D$_2$O & 470 mM & Extrusion\\
    Interferometry & 10.5$\pm$8.8 \cite{Betz:2012} & N/A & 295 mM Sucrose/300 mM Glucose & N/A & Electroformation\\
        
    \hline
    \centering
  \end{tabular*}
\end{table*}

\begin{table*}[h]
\small
  \caption{Bending rigidity values obtained in this study for DOPC under different conditions and microscopy settings. Note the sugar concentration is the same in all the experiments: 20 mM Sucrose inside/ 22 mM Glucose outside. The dye used is TR DHPE and all the vesicles were formed via electroformation.}
  \label{diffcondmethodhere}
  \begin{tabular*}{\textwidth}{@{\extracolsep{\fill}}llllll}
    \hline
    Microscopy & Rigidity ($\kT$) & Dye conc. ($\%mol$) & Objective/NA & Polarization Correction&  Pinhole (A.U)\\
    \hline
    Phase Contrast & 19.4$\pm$2.1 & 0.2 & 100x/1.25 & N/A & N/A\\
    Phase Contrast & 22.5$\pm$1.5 & 0 & 40x/0.6 & N/A & N/A\\
    Phase Contrast & 23.3$\pm$1.6 & 0.2 & 40x/0.6 & N/A & N/A\\
    Phase Contrast & 21.0$\pm$2.0 & 0.2 & 40x/1.3 & N/A & N/A\\
    Confocal & 21.7$\pm$2.0 & 0.2 & 40x/1.3 & Yes & 1\\
    Confocal & 22.5$\pm$2.1 & 0.2 & 40x/0.6 & Yes & 1\\
    Confocal & 22.5$\pm$2.4 & 0.2 & 40x/0.6 & Yes & 0.3\\
    Confocal & 22.6$\pm$3.5 & 0.2 & 40x/0.6 & Yes & 2\\
    Confocal & 20.4$\pm$4.0 & 0.2 & 40x/0.6 & No & 1\\
    Confocal & 22.3$\pm$1.6 & 0.2 & 40x/1.3 & No & 1\\
    Confocal & 25.0$\pm$2.1 & 2.0 & 40x/0.6 & Yes & 1\\
    \hline
    \centering
  \end{tabular*}
\end{table*}

 \section*{S2. Methods}
 
 \subsection*{Vesicle preparation}
 
Giant unilamellar vesicles (GUVs) were prepared using the classical electroformation method \cite{angelova.1987} from DOPC and the fluorescent lipid Texas Red 1,2-hexadecanoyl-sn-glycero-3-phosphoethanolamine (TR-DHPE). The composition of the GUVs explored are  99.8 \% DOPC 0.2 \% TR-DHPE and 98 \% DOPC 2 \% TR-DHPE (mole fractions). Stock solutions of DOPC and TR-DHPE at 10 mg/ml and 1 mg/ml in chloroform were diluted to a final concentration of
4 mM for varying proportions. A small volume, 10 $\mu$l, of the
solution was spread on the conductive surface of two glass
slides coated with indium tin oxide (ITO) (Delta Technologies).
The glass slides were then stored under a vacuum for 1--2 hours
to remove traces of organic solvent. Afterwards, a 2 mm Teflon
spacer was sandwiched between the glass slides and the
chamber was gently filled with 20 mM sucrose solution.
The slides (conductive side facing inward) were connected to an
AC signal generator Agilent 33220A (Agilent Technology GmbH,
Germany). An AC field of voltage 1.5 V and frequency 10 Hz
applied for 2 hours at room temperature, resulting in 10-50 $\mu$m
sized vesicles. The harvested vesicles were diluted 10 times in 22 mM glucose solution to obtain fluctuating vesicles. All GUVs were analyzed within 8 hours of electoformation.
 
 \subsection*{Microscopy and video recording}
 
 The equatorial fluctuations for both phase contrast and confocal mode were recorded with Leica TCS SP8 scanning confocal microscope using a HCX PL APO 40x/ Numerical Aperture (NA) 0.6 Ph2 (air) objective and a HC PL APO 40x/ NA 1.3 (oil) objective. The pinhole size during the experiment was fixed to 1 AU (Airy units) unless stated otherwise. Table 1 compiles the pixel size and focal depth for different experimental conditions. The scanning speed was fixed to 1 kHz in bidirectional mode and the polarizer plates were rotated (100\%) to remove the polarization effect of the fluorescent dye unless stated otherwise. The dye was excited with a 561 nm laser (diode-pumped solid-state laser) with \textcolor{black}{1.61\% (laser intensity)  HyD3 detector} (hybrid) and the gain was fixed to 23$\%$. Phase contrast imaging was recorded with PCO CS dimax (PCO AG,
Kelheim, Germany)) mounted on confocal microscope. 1500-2000 images were recorded at 3.83 frames per second (fps) with confocal and 60 fps  with phase contrast imaging. The RGB confocal images were converted to 8 bit and then inverted. We implemented an inbuilt MATLAB sobel disk filter $fspecial('sobel')$ and image normalization to increase the contrast of the contour.

 In this section, we list different focal depths and pixel sizes for different microscopy and numerical aperture settings for 40x objective. Focal depth or FWHM (full width half maximum) of phase contrast imaging was determined using the standard formula $d=\frac{\lambda}{NA^2}$ The wavelength of transmission light was assumed to be 550 nm.

\begin{table*}[h]
\small
  \caption{Different experimental conditions for video recording with 40x objective. }
  \label{tbl:example2}
  \begin{tabular*}{\textwidth}{@{\extracolsep{\fill}}lllllll}
    \hline
    Microscopy & Numerical Aperture & Medium & Pinhole size (AU) & Focal depth ($\mu${m}) & Pixel Resolution (nm)\\
    \hline
    Phase contrast & 0.6 & Air & 1 & 1.57 & 276.9\\
    Phase contrast & 1.3 & Oil & 1 & 0.35 & 158.7 \\
    Confocal & 0.6 & Air & 1 & 1.61 & 252.7\\
    Confocal & 1.3 & Oil & 1 & 0.52 & 252.7 \\
    \hline
  \end{tabular*}
\end{table*}

\begin{table*}[h]
\small
  \caption{Focal depth or FWHM (full width half maximum) for confocal imaging. }
  \label{tbl:example2}
  \begin{tabular*}{\textwidth}{@{\extracolsep{\fill}}lll}
    \hline
     Medium & Pinhole size (AU) & Focal depth ($\mu${m})\\
    \hline
    Air & 0.3 & 0.9\\
    Air & 1 & 1.6\\
    Air & 2 & 2.9\\
    \hline
  \end{tabular*}
\end{table*}

\subsection*{Sub-pixel contour recognition}
\begin{figure}[h]
\includegraphics[scale=.5]{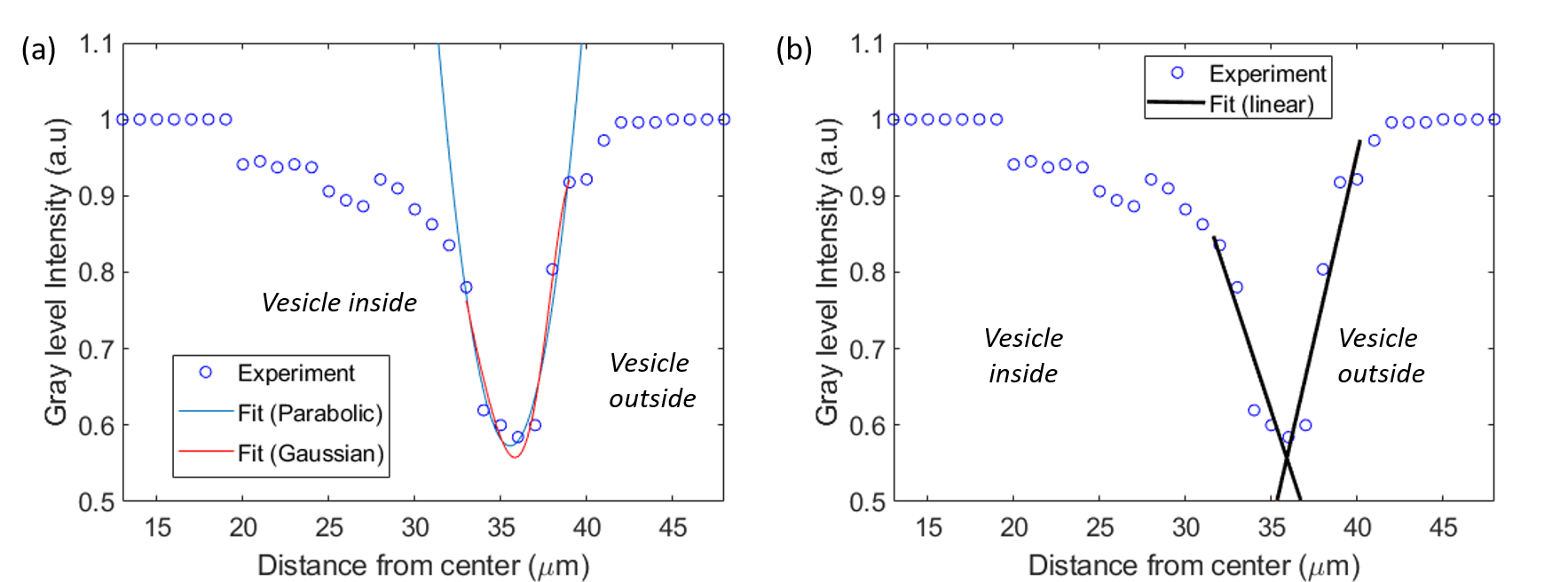}
\centering
\caption{Intensity profile for a vesicle contour obtained from confocal imaging. The contour recognition details are given in \cite{gracia.2010}. The sub-pixel accuracy of the contour profile is determined based on (a) Gaussian, parabolic and (b) linear interpolations}
\centering
\label{fig2_subpixel}
\end{figure}

The intensity profile  in the radial direction for N wedges were determined from three different interpolation schemes (Gaussian, parabolic and linear weighting of neighbouring pixel) for sub-pixel contour recognition. This was done to check if different interpolation schemes affects the bending rigidity values due to uncertainty introduced at higher wave-numbers for experimental vesicle contour fluctuations. The mean bending rigidity obtained was similar for all the three schemes for the same vesicle. \reffig{fig2_subpixel} illustrates the subpixel accuracy determination for a 35 $\mu$m radius vesicle. The bending rigidities obtained was 22.0$\pm$3.0 $\kT$, 21.1$\pm$1.0 $\kT$ and 21.9$\pm$2.2 $\kT$ from Gaussian, parabolic and linear interpolation schemes respectively.

\section*{S3. Polarization Effects}

We analyzed the same vesicle with and without polarization effects. The polarization effects were corrected using circular plates that were rotated 100$\%$. \reffig{figpol} illustrates the effect of dye polarization for vesicles imaged with different numerical apertures. Using one Anova test, we find a significant difference of 3 $\kT$ for the 40x/0.6 NA case. The difference tends to be negligible for 40x/1.3 NA case. 

\begin{figure}[h]
\includegraphics[width=\textwidth]{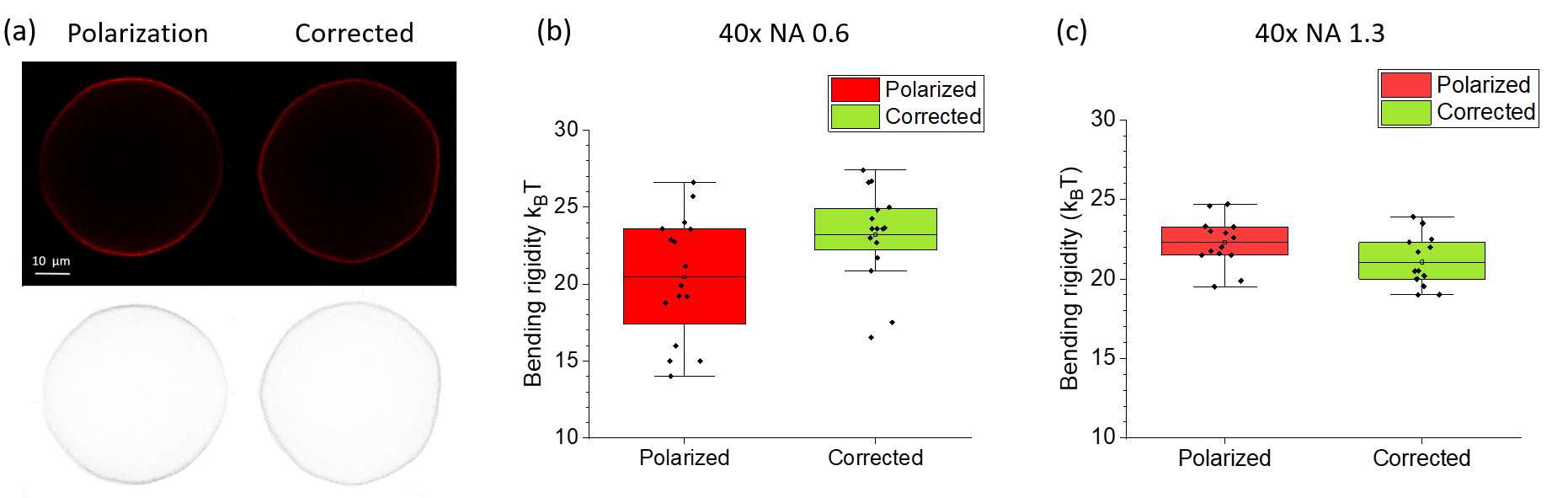}
\centering
\caption{Polarization effects. (a) Confocal images of the same vesicle with and without polarization effects for 40x/0.6 NA case. The polarization effects were removed using circular plates that were rotated 100$\%$. (b, c) Comparison between the same vesicles for different numerical apertures. Using one Anova test, we find a significant difference of 3 $\kT$ for 40x/0.6 NA case. The difference tends to be negligible for 40x/1.3 NA case. Pinhole size is 1 AU. }
\centering
\label{figpol}
\end{figure}

\section*{S4. Effect of Vesicle Size on Bending Rigidity Values}

The bending rigidity obtained from confocal microscopy with low-resolution optics (e.g. 40x objective, NA 0.6, 1 AU, polarization correction) can be systematically underestimated if the vesicle population contains similar sized vesicles. We demonstrate this by comparing the bending rigidity of the same vesicle imaged with confocal and phase-contrast microscopy, see \reffig{figbrsize}. Vesicles with smaller sizes yield apparently lower bending rigidity, see Fig. S4 which further highlights the bias effect. For small vesicles, the out-of-focus signal gives rise to asymmetry in the contour intensity (illustrated in Fig. 4a  in the main text) which leads to errors in the contour detection and underestimation of the bending rigidity. When the refractive index difference across the membrane is small (as is the case in our experiments), phase contrast imaging does not suffer from this size bias. 

\begin{figure}[h]
\includegraphics[scale=.3]{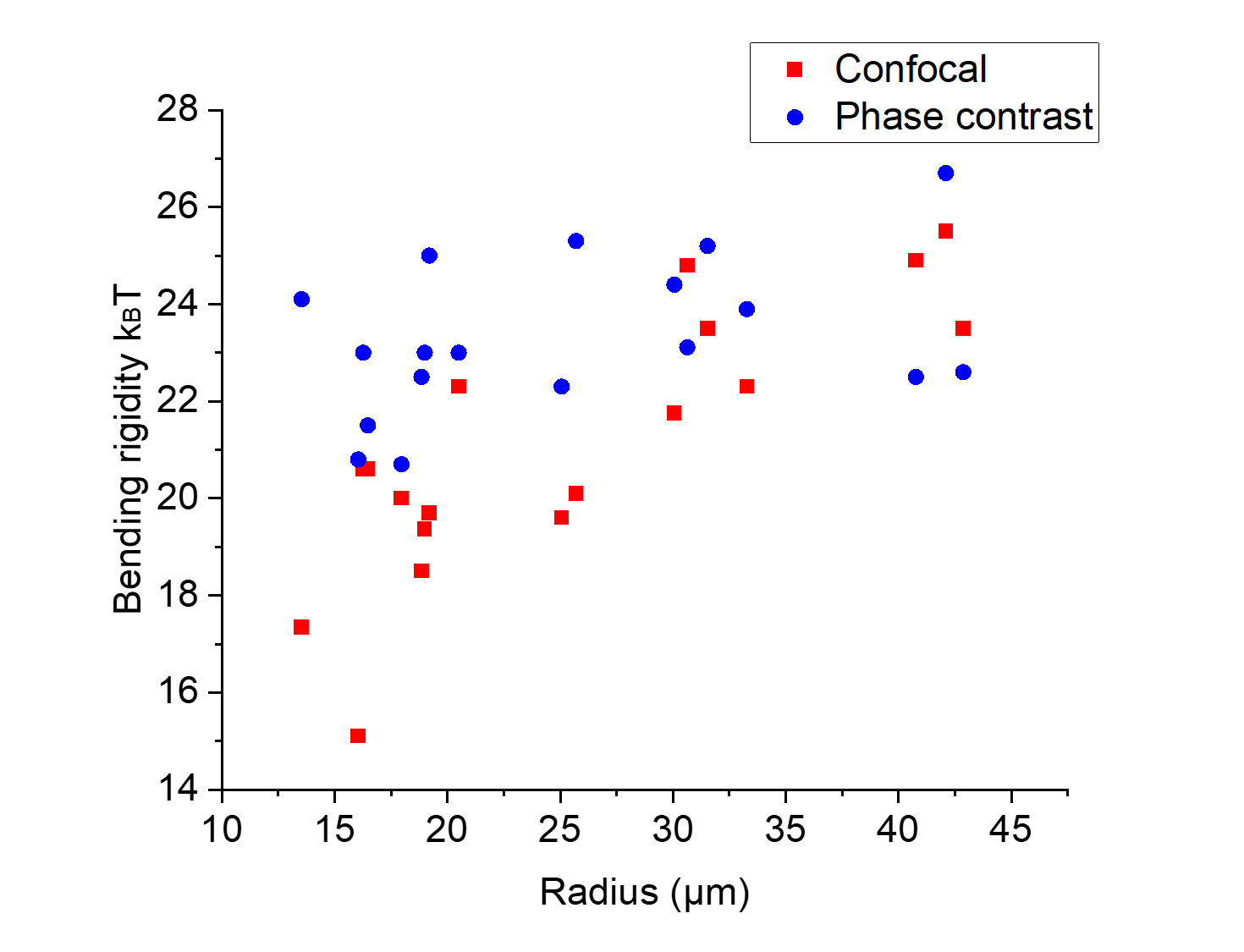}
\centering
\caption{Vesicle size effects. Every vesicle was imaged with confocal and phase contrast microscopy. Data are collected on DOPC vesicles with different sizes. The dye concentration was 0.2
mol. Imaging was done with 40x objectives with NA 0.6, 1 AU and polarization correction.  }
\centering
\label{figbrsize}
\end{figure}

\section*{S5. Bootstrapping resampling}
\begin{figure}[h]
\includegraphics[scale=0.65]{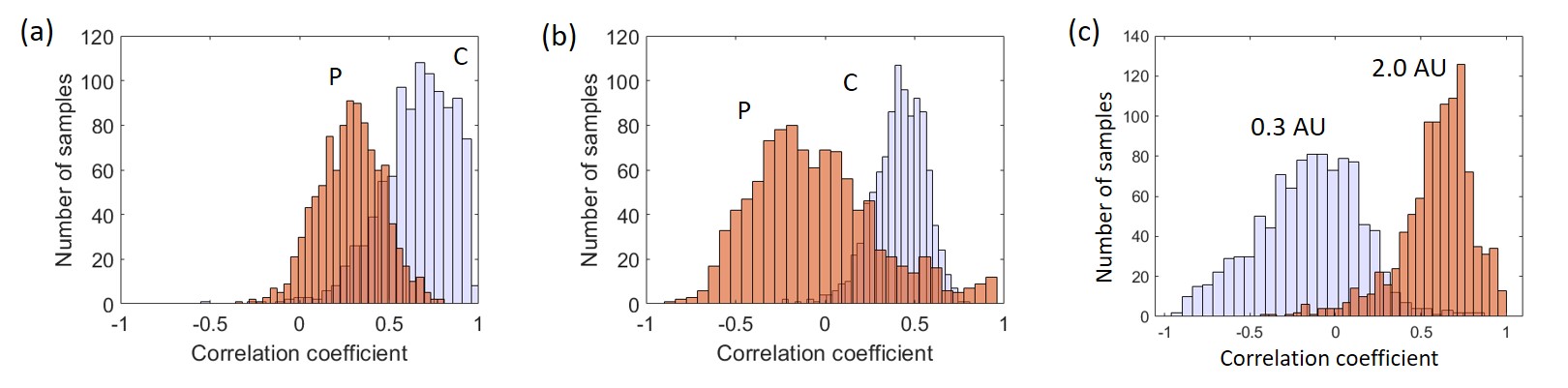}
\centering
\caption{ Bootstrap method with 95$\%$ confidence to evaluate  bending rigidity dependence on size of vesicles for different numerical aperture (a) 40x/0.6 NA, (b) 40x/1.3 NA in phase contrast (P) or confocal(C) microcopy, and the pinhole sizes (c) (blue AU 0.3 and  red AU 2).  }
\centering
\label{figBSl}
\end{figure}

Details about the various statistical techniques can be found in Ref. \cite{maxwell.2017}. Here we explain the bootstrapping sampling technique. A more rigorous reference is the textbook \cite{chernick.2007}. In practice, the finite amount of data or length of experiment limits the accuracy to infer data confidently. Bootstrapping is an inference method about the population from a given sample. In bootstrap-resamples, the population is in fact the sample and this quantity is known. This allows to measure the quality of inference of the 'true' sample from a re-sampled data. For example, let's consider the average mass of the human population world wide. It is difficult to measure the mass of every individual globally, therefore, a small sample is measured. Let's assume the sample size of N people. From that sample size, only one mean can be measured. In order to have a reasonable estimate about the population statistics, we need to have variability of the mean that we computed. The simplest bootsampling statistics can be considered by taking the original data N individuals and resampling to create a new sample of the same size N (e.g. we might 'resample' 10 times from [60,61,62,63,64,65,66,67] kg and get [61,64,63,63,60,60,62,65] kg). This process is repeated a large number of times, 100 to 10000, to create a histogram that be applied to any estimator testing.  Bootstrap resampling was carried out using MATLAB's \textit{bootstrp ()}.

In the case of our experiments, the finite amount of data or length of experiment limits the accuracy to infer data confidently. The bootstrap resampling requires choosing random replacement from a given data set and examining each sample the same way. This way a particular data point from the original set can reappear randomly multiple times in a particular bootstrap sample. The element size of the bootstrap sample is the same as the element size of the original data. This technique allows to obtain uncertainty of the quantity one estimates.  

Bootstrap resampling algorithm for estimating standard error \cite{chernick.2007}: \newline
1. Obtain N independent bootstrap samples $X^{*1}, X^{*2}, X^{*3}, ...X^{*N}$, each consisting of n data values drawn with a replacement from $x$ where $x=[x^1,x^2,x^3...x^n]$. Note for estimating a standard error, the number N will ordinarily be larger than 30 to satisfy the Central Limit Theorem. Computations allow to use a large number N such as $10^3$ to $10^4$.\newline
2. Determine the bootstrap replication for every bootstrap resample:

\begin{align}
     {\zeta^*(b)}=s(X^{*b}) \,\,\,\,\,\,\,\,\, b=[1,2,3,..N]
\end{align}

where $s()$ is a statistical function like sample mean. For example, if $s(x)$ is the sample mean $\Bar{x}$ then $S(X^*)$ is the mean of bootstrap data set. \newline
3. Compute the standard error $SE$ by utilizing the standard deviation of N replications

\begin{equation}
    SE=\bigg[\frac{{\sum_{b=1}^N}[{\zeta^*(b)}-{\zeta^*(.)}]}{N-1}\bigg]^\frac{1}{2}
\end{equation}

where ${\zeta^*(.)}=\sum_{b=1}^N\zeta^*(b)/N$.\newline

In our case we determine the $SE$ of mean Pearson correlation using bootsampling statistics.

\section*{S6. Numerical simulations of vesicle contours}

\subsection*{Mathematical Model}

The total energy of the system is given by the Helfrich model\cite{Seifert:1997} as Eq. (\ref{Eq-1}) where $\kappa$ is the bending rigidity, $c_1$ and $c_2$ are the local radii curvatures,  $A$ is the total surface area, $V$ is the interior volume of the vesicle, $\sigma$ is the surface tension, and $p$ is the pressure difference across the membrane.
\begin{equation}
    F = \frac{\kappa}{2}\int _A ( c_1 + c_2)^2 dA + \sigma A + pV
    \label{Eq-1}
\end{equation}
For a quasi-spherical vesicle in equilibrium, the shape can be decomposed into spherical harmonics ($\mathcal{Y}_{lm}$)  such that the position of the surface is given by 
\begin{equation}
    R(\theta, \phi, t) = R_0 \Bigg( 1 + \sum_{l= 0}^{l_{max}} \sum_{m=-l}^l f_{lm}(t) \mathcal{Y}_{lm} (\theta, \phi)\Bigg)
    \label{Eq-2}
\end{equation}
where the characteristic radius $R_0$ is given by $V = \frac{4}{3} \pi R_0^3$. The spherical harmonics are defined as
\begin{equation}
\label{spherY}
\mathcal{Y}_{lm}=n_{lm} P_{lm}(\cos\theta) e^{\im m \phi}\,,\quad n_{lm}=\sqrt{\frac{(2l+1)(l-m)!}{4\pi (l+m)!}}
\end{equation}
$P_{lm}(\cos\theta) $ are the associated Legendre polynomials.

As $l=1$ account for translational modes, for the sake of this paper, $f_{lm}(t)$ will be  restricted to $f_{1m}(t)=0$ for $l=1$. Furthermore, volume conservation requires that \cite{Seifert:1999} 
\begin{equation}
     f_{00} =  \frac{-1}{\sqrt{4\pi}}\sum_{l=2}^{l_{max}} \sum_{m=-l}^l |f_{lm}|^2.
     \label{Eq-3}
\end{equation}
\begin{figure}[h]
\includegraphics[scale=.55]{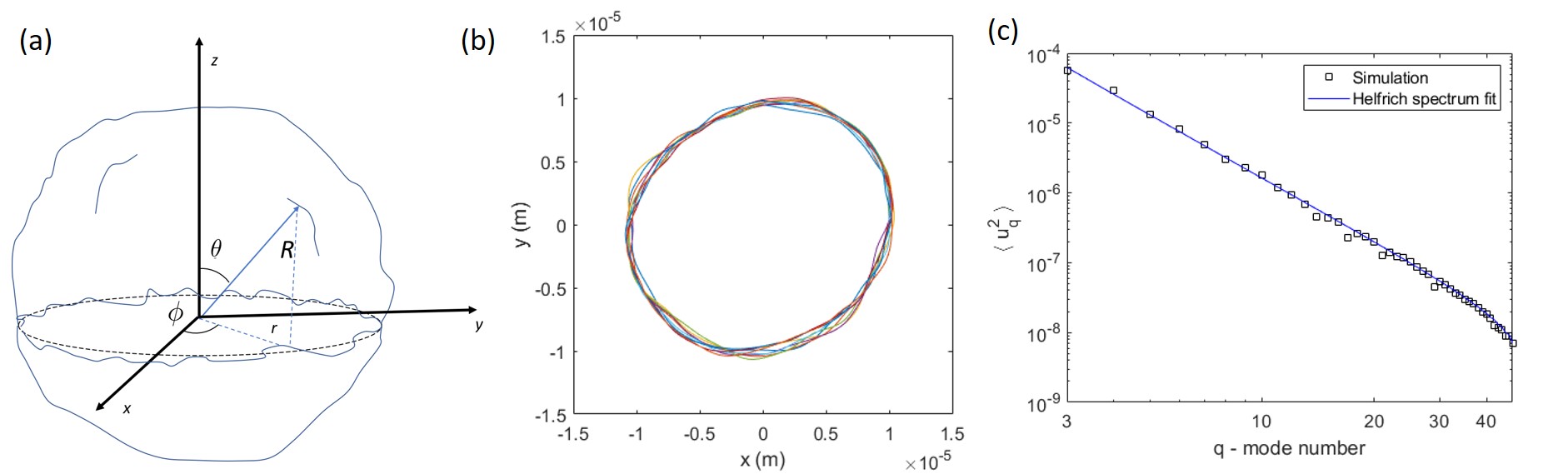}
\centering
\caption{(a) A  sketch of a GUV. (b) Time sequence of vesicle contours taken at time intervals of 1 s; the bending rigidity  is $\kappa=$1$\times\,10^{-19}$ J and the membrane tension is $\sigma=$1$\times\,10^{-9}$ N/m. The size of the vesicle is $R_0=10^{-5}$ m. (c) Helfrich mode spectrum determined by the image detecting algorithm based on Ref. \cite{gracia.2010}. The spectrum was fitted with Equation 2 from the main text to obtain the bending rigidity and membrane tension. }
\centering
\label{figS}
\end{figure}
Assuming there is no external fluid flow, the harmonic coefficients ($f_{lm}$) for $l>1$ are described by the following stochastic  differential equation \col{\cite{Seifert:1999}}
\begin{equation}
    \partial_t f_{lm} = - \tau_l^{-1} f_{lm} + \zeta_{lm}(t)
    \label{Eq-4}
\end{equation}
where 
\begin{equation}
\label{taul}
 \tau_l = \frac{\eta_\out R_0^3}{\kappa \Gamma_l (\lambda)E_l}, \quad  \Gamma_l = \frac{l(l+1)}{4l^3 + 6l^2 -1+\left(2l^3+3l^2-5\right)\left(\lambda-1\right)} \quad \text{and} \quad E_l = (l+2)(l-1) \Big( l(l+1) + \bar \sigma \Big).
\end{equation}
A tutorial derivation of the evolution equation and the  relaxation time (in the absence of thermal noise) can be found in Refs. \cite{Vlahovska:Stone,Vlahovska:2019book}
The dimensionless tension is $\bar\sigma=\sigma R_0^2/\kappa$.
$\lambda={\eta_\ins}/{\eta_\out}$ is the ratio of viscosities of the solutions inside and outside the vesicle. When $\lambda=1$, our result for the relaxation time reduces to the one reported by Refs. \cite{Milner-Safran:1987,Seifert:1999}. To make  easier comparison with the result of Ref. \cite{Rautu:2017}, we can rewrite the relaxation time as
\begin{equation}
\label{taul}
 \tau_l = \frac{R_0^3}{\kappa}  \frac{\eta_\out\left(2l^3 + 3l^2 +4\right)+\eta_\ins\left(2l^3+3l^2-5\right)}{l(l+1)(l+2)(l-1) \Big( l(l+1) + \bar \sigma \Big)}
\end{equation}

$\zeta_{lm}(t)$ is a stochastic term accounting for thermal noise; the corresponding time correlation is given as 
\begin{equation}
    \Big<\zeta_{lm}(t) \zeta_{l'm'}(t')\Big> = (-1)^{m} \ \frac{2 k_B T \Gamma_l}{\eta_\out R_0^3} \ \delta_{l,l'} \delta_{m,-m'} \delta(t-t').
    \label{Eq-5}
\end{equation}
The $\delta$ functions are the traditional Kronecker and Dirac delta functions. From Eq. \ref{Eq-5}, the variance of $\zeta_{lm}(t)$ is given by
\begin{equation}
    \Big<|\zeta_{lm}|^2 \Big> = 2\frac{ k_B T \Gamma_l}{\eta_\out R_0^3} = 2 \Sigma_l.
    \label{Eq-6}
\end{equation}

\subsection*{Numerical Method}

At this point, it is convenient to decompose $f_{lm}$ and $\zeta_{lm}$ into real and imaginary components such that $f_{lm}(t) = X_{lm}(t)  + \im \ Y_{lm}(t)$ and $\zeta_{lm}(t) = a_{lm}(t)  + \im \ b_{lm}(t)$. As $a_{lm}$ and $b_{lm}$ are independent of each other then 

\begin{equation}
    \Big<|\zeta_{lm}|^2 \Big> = \Big<|a_{lm}|^2 \Big> + \Big<|b_{lm}|^2 \Big> = 2 \Big<|a_{lm}|^2 \Big> = 2 \Big<|b_{lm}|^2 \Big> = 2 \Sigma_l
    \label{Eq-7}
\end{equation}
Eq. (\ref{Eq-4}) can then be rewritten as 
\begin{equation}
    \partial_t X_{lm} = - \tau_l^{-1} X_{lm} + a_{lm}(t)
    \label{Eq-8}
\end{equation}
and similarly for $Y_{lm}$.
 As Eq. (\ref{Eq-8}) is a simple Langevin equation, the exact time update \cite{Gillespie: 1996} is given as 
  \begin{equation}
        X_{lm}(t+\Delta t) = X_{lm}(t) e^{-\Delta t/\tau_l} + \bigg[ \frac{1}{2} \Sigma_l^2 \tau_l \big(1 - e^{-2\Delta t /\tau_l}\big)\bigg]^{1/2} n
     \label{Eq-9}
 \end{equation}
such that $\Delta t$ is the time step size and $n$ is a sample value from the normal distribution $\mathcal{N}(0,1)$. In order to properly resolve the dynamics of the higher order coefficient, a sufficiently small time step must be chosen so that $\Delta t << \tau_{l_{\max}}$. Yet as each harmonic coefficient is independent of each other, Eq. \ref{Eq-9} can be evaluated for all $X_{lm}$ and $Y_{lm}$ simultaneously. Given all the harmonic coefficients ($f_{lm}$), the cross-section at the equator, $R(\theta= \pi/2)$,  can easily be computed using Eq. (\ref{Eq-2}).

When running the numerical simulations, the user has some choice of which input parameters  to specify. 
For example, one can specify the effective surface tension (dimensionless) $\bar\sigma$ and the largest incorporated mode $\l_{max}$. In this case, the vesicle's excess is obtained from \cite{Seifert:1999}
\begin{equation}
    \alpha = \frac{\kT}{2\kappa} \Bigg[ \frac{5}{6+\sigma} + \ln\bigg( \frac{l_{max}^2 +\bar\sigma}{12+\bar\sigma} \bigg) \Bigg],
    \label{lmaxEq1}
\end{equation}
Alternatively, one can  specify $\alpha$ and $\bar \sigma$, and \refeq{lmaxEq1} then provides the requisite $l_{max}$.

Here we demonstrate an example of a numerically simulated vesicle with predefined bending rigidity and membrane tension. \reffig{figS}b shows a time sequence of equatorial vesicle contours with bending rigidity of $\kappa=10^{-19}$ J and membrane tension of $\sigma=10^{-9}$ N/m. The size of the vesicle is $R_0=10^{-5}$ m. By implementing our image detection technique and fitting algorithm from \textit{Gracia et al.}\cite{gracia.2010}, we are able to reproduce the bending rigidity and membrane tension respectively as  $\kappa=(1.00 \pm\,0.01)\times\,10^{-19}$ J and $\sigma=(1.1 \pm\,0.2)\times\,10^{-9}$ N/m with the Helfrich spectrum given in \reffig{figS}. Notably our image detection is able to resolve more than 45 shape fluctuation modes.

\subsection*{Simulating the Effect of Out-of-focus Signal}

Due to a finite focal depth, the microscope imaging does not capture only the optical/fluorescence signal at the focal (equatorial) plane. 
The out-of-focus signal results in gradient in the image intensity near the focal plane vesicle contour. 

To simulate this effect, we numerically projected the vesicle shape  $R(\theta,\phi,t)$ on the equatorial plane and assigned intensity of the projected location, $R(\theta,\phi,t)\sin\theta$, given by
\begin{equation}
     I(r,\phi,t) = \int_{0}^{2\pi} \int_{\frac{\pi}{2}-\theta_{fd}}^{\frac{\pi}{2}+\theta_{fd}} W(\theta') \ \delta(\phi-\phi') \ \delta(r - R(\theta',\phi',t) \sin\theta') \ d\theta' d\phi',
     \label{EqVesc-14}
 \end{equation} 
where $\frac{\pi}{2}\pm\theta_{fd}$ are the top and bottom of the microscope focal depth ($FD$), $\theta_{fd}=\arctan(FD/R_0)$. $W(\theta)$ is the intensity weighting function
\begin{equation}
     W(\theta) = \frac{1}{W_0} \exp\bigg[-\frac{\cos^2(\theta)}{2\cos^2(\theta_{fd})}\bigg]
     \label{EqVesc-14a}
 \end{equation}
 and $W_0$ is the corresponding normalization constant. The resulting images of the equatorial plane at different focal depth are illustrated in Figure S\ref{FIG_sim}a.

 We varied the magnitude of the focal depth $FD$, from 0 to 0.3$R_0$. 
  The fluctuation spectra obtained for the simulations are shown in Figure S\ref{FIG_sim}b for a vesicle sized $R_0=20\,\mu{m}$ with  $\kappa$= 22 $\kT$ and $\sigma= 1.4\times$10$^{-9}$ N/m. The crossover mode $q_c=\sqrt{\bar\sigma}\sim3$. The effect of the projections is only significant for modes $q\geq{\Delta}^{-1}$, where  $\Delta=\frac{FD}{R_0}$ \cite{Rautu:2017}. For smaller values of $\Delta<0.05$, the projections have no effect - the spectra overlap  implying same bending rigidity.   However, as the value of $\Delta$  increases,
  more modes get affected by the projections resulting in an effective softening of the membrane from 22 $\kT$ to 19 $\kT$, see  S\ref{FIG_sim}c.


\begin{figure}[h]
\includegraphics[width=\linewidth]{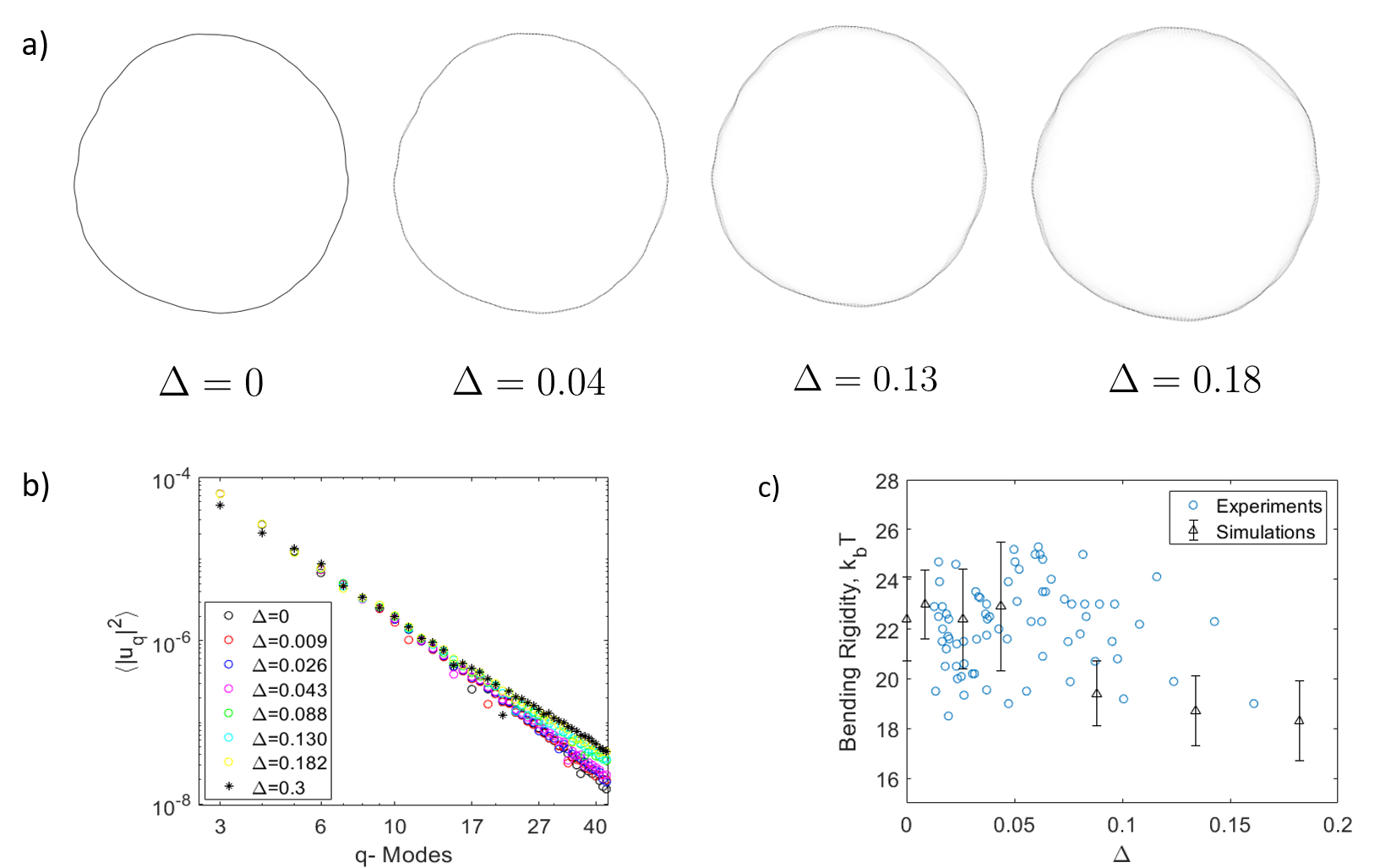}
\centering
\caption{a) Snapshot of vesicle equatorial contours at different $\Delta=FD/R_0$. The simulated vesicle has  bending rigidity= 22 $\kT$, membrane tension= 1.4$\times$10$^{-9}$ N/m and radius  $R_0=20\,\mu{m}$. Each image was acquired  over 0.2 s (corresponding to imaging rate of 5 fps). b) Fluctuation spectrum obtained at different $\Delta$ from the numerical simulations c) Bending rigidity obtained for different $\Delta$. Here we have compared the experimental results with numerical simulations.}
\centering
\label{FIG_sim}
\end{figure}

\subsection*{Fluctuations statistics: derivations of the basic results}
Here we summarize the main results for the dynamics of a quasi-spherical vesicle.

\paragraph*{Mean Squared Magnitude of the Fourier Modes:}

The dynamics of the spherical harmonics modes is governed by the following Langevin equation,
\begin{equation}
    \frac{\partial f_{lm}}{\partial t} = - \tau_l^{-1} \ f_{lm} +\zeta_{lm}
    \label{Eq-LA1}
\end{equation}
where the relaxation time  $\tau_l$ is given by \refeq{taul} and the noise is
\begin{equation}
  \Big< \zeta_{lm} \Big>=0 \text{   and   }  \Big< \zeta_{lm}^*(t) \zeta_{l'm'}(t') \Big> = \frac{2\kT\Gamma_l}{\eta_{\text{ex}} R_0^3} \delta(t-t') \delta_{ll'} \delta_{mm'}.
    \label{Eq-LA2}
\end{equation}
The analytic solution to Eq. (\ref{Eq-LA1}) is given by
\begin{equation}
    f_{lm}(t) = e^{-t/\tau_l} f_{lm}(0) + \int_0^t e^{-(t-t')/\tau_l} \zeta_{lm} (t') dt'.
    \label{Eq-LA3}
\end{equation}
and therefore
\begin{multline}
    |f_{lm}(t)|^2 = e^{-2t/\tau_l} |f_{lm}(0)|^2 + \int_0^t e^{-(2t-t')/\tau_l} \Big( f_{lm}(0) \zeta_{lm}^* (t') + f_{lm}(0)^* \zeta_{lm} (t') \Big) dt' \\ + \int_0^t \int_0^t e^{-(2t-t'-t'')/\tau_l} \zeta_{lm} (t') \zeta_{lm}^* (t'') dt' dt''.
    \label{Eq-LA4}
\end{multline}
The ensemble average of $\Big< |f_{lm}|^2 \Big>$ of Eq. (\ref{Eq-LA4}) is then 
\begin{multline}
    \Big<|f_{lm}|^2\Big>=  e^{-2t/\tau_l} |f_{lm}(0)|^2+ \int_0^t e^{-(2t-t')/\tau_l} \Bigg( f_{lm}(0) \Big<\zeta_{lm}^* (t')\Big> + f_{lm}(0)^* \Big< \zeta_{lm} (t')\Big> \Bigg) dt'  \\ + \int_0^t \int_0^t e^{-(2t-t'-t'')/\tau_l} \Big<\zeta_{lm} (t') \zeta_{lm}^* (t'')\Big> dt' dt''.
    \label{Eq-LA5}
\end{multline}
Using Eq. (\ref{Eq-LA2}), Eq. (\ref{Eq-LA5}) simplifies to 
\begin{equation}
    \Big<|f_{lm}|^2\Big>=  e^{-2t/\tau_l} |f_{lm}(0)|^2+  \frac{2\kT \Gamma_l}{\eta_{\text{ex}} R_0^3} \int_0^t e^{-2(t-t')/\tau_l}  dt'.
    \label{Eq-LA6}
\end{equation}
\begin{equation}
    \Big<|f_{lm}|^2\Big>=  e^{-2t/\tau_l} |f_{lm}(0)|^2+  \frac{\kT \Gamma_l \tau_l}{\eta_{\text{ex}} R_0^3} \Big[1 - e^{-2t/\tau_l}\Big]
    \label{Eq-LA7}
\end{equation}
At long times, $t>>\tau_l$, Eq. (\ref{Eq-LA7}) simplifies to 
\begin{equation}
    \Big<|f_{lm}|^2\Big>=  \frac{\kT \Gamma_l \tau_l}{\eta_{\text{ex}} R_0^3} = \frac{k_B T}{\kappa} \bigg[ (l+2)(l-1)\Big( l(l+1) + \bar\sigma \Big) \bigg]^{-1}
    \label{Eq-LA8}
\end{equation}
Recall $\bar\sigma=\sigma R_0^2/\kappa$.
Since the dynamics of the different spherical harmonics modes are completely decoupled, we can more generally say 
\begin{equation}
    \Big<f_{lm}^* f_{l'm'}\Big>=  \frac{k_B T \Gamma_l \tau_l}{\eta_{\text{ex}} R_0^3} =  \frac{k_B T}{\kappa}\bigg[ (l+2)(l-1)\Big( l(l+1) \kappa + \bar\sigma  \Big) \bigg]^{-1} \delta_{ll'} \delta_{mm'}
    \label{Eq-LA8a}
\end{equation}
Next, we consider the contour of the GUV at the equator as a function of the spherical harmonic coefficients:
\begin{equation}
    r(\phi,t) =  R_0 \left(1+ \sum_{q=0}^{q_{\max}} u_q(t)  e^{\im q \phi}  \right)=R_0 \Big(1+ \sum_{l=0}^{l_{\max}} \sum_{m=-l}^{l} f_{lm}(t) {\mathcal{Y}}_{lm} (\pi/2,\phi) \Big).
    \label{Eq-LA10}
\end{equation}
The Fourier coefficient for the $q$-th mode is then given by 
\begin{equation}
    u_q(t) = \frac{1}{2\pi R_0}\int_0^{2\pi} r(\phi,t)  e^{-iq\phi} d\phi = \sum_{l=q}^{l_{\max}} f_{lq}(t) \left(n_{lq}P_{lq} (0) e^{\im q\phi}\right) e^{-\im q\phi}
    \label{Eq-LA11}
\end{equation}
as all the other terms integrate to zero. In the above equation, we have inserted the definition of the spherical harmonic, ${\mathcal{Y}}(\theta, \phi)=n_{lm}P_{lm}(\cos\theta) e^{\im m \phi}$ (see \refeq{spherY}), which shows that the dependence on $\phi$ cancels out.

The mean squared amplitude of $u_q$ is then given by 
\begin{equation}
    \big<|u_q|^2\big> =  \sum_{l=q}^{l_{\max}} \sum_{l'=q}^{l_{\max}} \big<f_{l'q}^* f_{lq}\big> n_{lq}n_{l'q}P_{lq} (0) P_{l'q}^* (0).
    \label{Eq-LA12}
\end{equation}
Using Eq. (\ref{Eq-LA8a}), the above equation simplifies to 
\begin{equation}
 \big<|u_q|^2\big> =  \sum_{l=q}^{l_{\max}} \big<| f_{lq}|^2\big>n_{lq}^2 |P_{lq} (0)|^2
\label{Eq-LA13}
\end{equation}
 \begin{equation}
 \big<|u_q|^2\big> =  \kT \sum_{l=q}^{l_{\max}}  \bigg[ (l+2)(l-1)\Big( l(l+1) \kappa+ \sigma  R_0^2 \Big) \bigg]^{-1} n_{lq}^2|P_{lq} (0)|^2
\label{Eq-LA14}
\end{equation}
\refeq{Eq-LA14} follows $q^{-3}$ behavior for bending dominated modes $q>\sqrt{\bar\sigma}$ (and $q^{-1}$ behavior for tension dominated modes $q<\sqrt{\bar\sigma}$). 


\vspace{1cm}

\paragraph*{Time Correlation for Fourier Modes:} Time correlations present another useful metric to analyze the membrane fluctuations.
 As the different spherical harmonics modes are independent, the average time correlations,
  \begin{equation}
     \Big< u_q (0) u_q^*(t) \Big> =  \sum_{l'=|q|}^{l_{max}} \sum_{l''=|q|}^{l_{max}} \Big<  f_{l'q}(0) \ f_{l''q}^*(t) \Big> n_{l'q} n_{l''q}P_{l'q} (0) \ P_{l''q}^* (0),
     \label{Eq7}
 \end{equation}
 can be simplified to 
  \begin{equation}
     \Big< u_q (0) u_q^*(t) \Big> =  \sum_{l=|q|}^{l_{max}} \Big<  f_{lq}(0) \ f_{lq}^*(t) \Big>n_{lq}^2 \Big|P_{lq}(0)\Big|^2 \ ,
     \label{Eq8}
 \end{equation}
 Using \eqref{Eq-LA4}, \eqref{Eq7} can be rewritten as
 \begin{equation}
     \Big< u_q (0) u_q^*(t) \Big> = \sum_{l=|q|}^{l_{max}} \Big<  |f_{lq}|^2 \Big> n_{lq}^2 \Big|P_{lq}(0)\Big|^2 e^{-t/\tau_l}.
     \label{Eq9}
 \end{equation}
 Since the first term in \eqref{Eq9} has both the smallest decay rate ($\tau_q^{-1}$) and largest mean-squared amplitude, the time correlation can be approximated to leading order as 
  \begin{equation}
     \Big< u_q (0) u_q^*(t) \Big> =  \Big<  |f_{qq}|^2 \Big>  n_{qq}^2 \Big|P_{qq} (0)\Big|^2 e^{-t/\tau_q}.
     \label{Eq10}
 \end{equation}
 If we consider limit of undulations with short wavelengths (shorter than the vesicle radius), $q\gg1$, then the leading order decay rate can be approximated as  
 \begin{equation}
    \tau_q^{-1} \approx \frac{q^3 R_0^{-3} \kappa + q R_0^{-1} \sigma}{2( \eta_\out+\eta_\ins)}=
\frac{ \kappa }{\eta_\out R_0^3}\frac{q^3  + q \bar  \sigma}{2( 1+\lambda)}
    \label{Eq12}
\end{equation}
which is the decay rate derived using planar fluctuations. However, we suggest using the exact decay rate from the spherical harmonics as it is both more accurate and valid for all Fourier modes.

\begin{figure}[h!]
    \centering
        \includegraphics[height=2.6in]{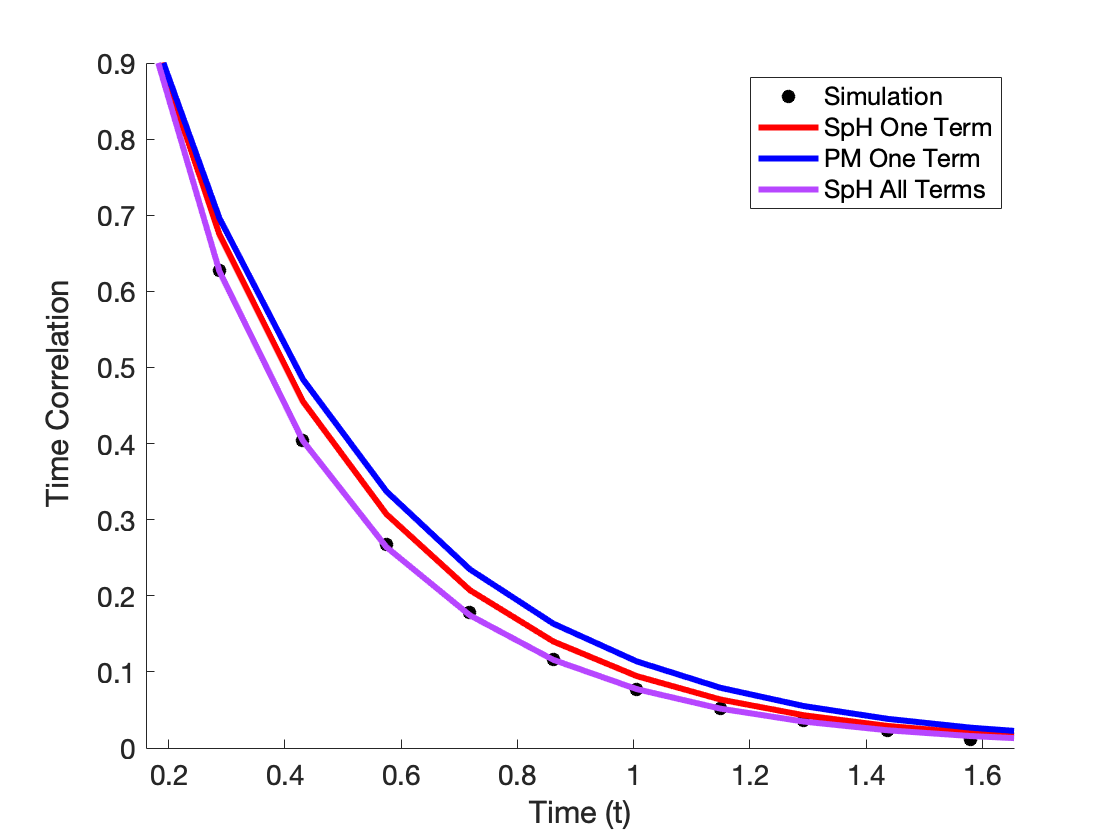}
        \includegraphics[height=2.6in]{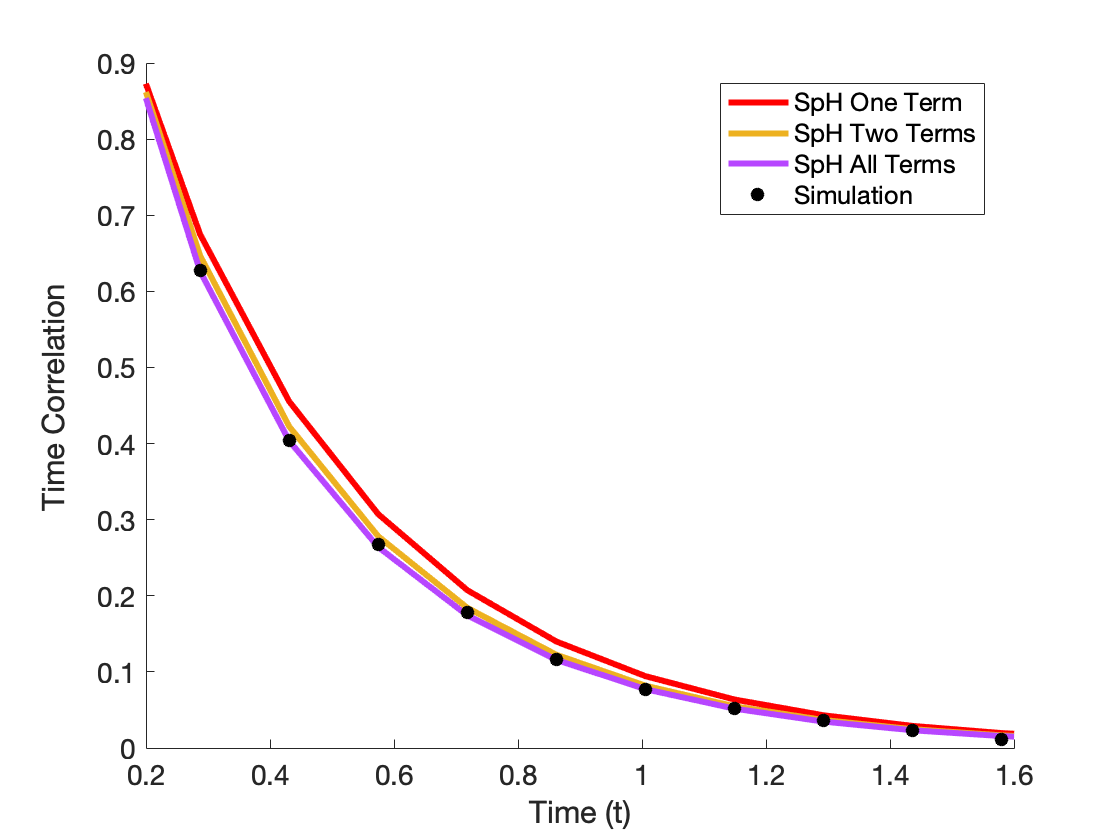}
    \caption{Plots comparing the analytic approximations for time correlation for Fourier mode $q=5$. The left plots the time correlations using the exact spherical harmonic (SpH) decay rate and the less accurate planar membrane (PM) decay rate. The right plots the time correlations for the SpH case using different number of terms. The black dots show the time correlations computed from a numerical simulation using the following parameters as inputs. $R_0 = 3\times 10^{-5}$ m, $\kappa = 5\times10^{-19}$ J, $\sigma=4\times10^{-8}$ N/m, $l_{max}=14$}
    \label{figTC}
\end{figure}

When comparing the time correlations in Fig.\ref{figTC}, the exact decay rate, from the full spherical harmonics (SpH), is immediately more accurate than if the planar membrane (PM) decay rate is used. To get the accuracy even better, the higher order terms in Eq. (\ref{Eq9}) must be included. If all of the terms are included then the time correlation is directly on top of the curve from produced by the numerical simulation. However, as it is not feasible to include all the terms for real membranes, it is of interest to know how many terms are enough to sufficiently reproduce the numerical simulations. As shown in Figure 5, the time correlation produced by including the first two terms in Eq. (\ref{Eq9} lies almost directly on top of the true solution. Including more terms would improve the accuracy further, but it is not likely to be significant due to experimental error.
\vspace{0.5cm}
\paragraph*{Cross-Spectral Density:}
Similar to time correlations, the Cross-Spectral Density (CSD) is given by
\begin{equation}
   \Big< |u_q (0)| |u_q(t)| \Big> - \Big< |u_q|^2 (0)\Big>.
   \label{Eq13}
\end{equation}
For the sake of clarity of explanation, in this section we will use the leading order approximation of $u_q$,
\begin{equation}
    u_q(t) 
    \approx  f_{qq}(t) n_{qq} P_q \left(\cos\frac{\pi}{2}\right) \,.
    \label{Eq14}
\end{equation}
Using \eqref{Eq-LA3}, this can be rewritten as 
\begin{equation}
    u_q(t) = e^{-t/\tau_l} u_q(0) + \Bar{\zeta}_q(t),
    \label{Eq15}
\end{equation}
where 
$$ \Bar{\zeta}_q = n_{qq}P_{qq} (0) \int_0^t e^{-(t-t')/\tau_l} \zeta_{lm} (t') dt'$$
is a random normally distributed Weiner process.
 
 From \eqref{Eq15}, it is clear that  $u_{q}(t)=\Bar{\zeta}_{q}(t)$ for large values of $t$. Furthermore, it is worth noting that all Fourier modes, except $q = 0$, have both real and an imaginary component, $u_q = A_q + \im B_q$, and that these two components are independent of each other. Likewise, the thermal noise can be decomposed into independent real and imaginary components:  $\Bar{\zeta}_{q}=\Bar{\zeta}_{Aq}+\im \Bar{\zeta}_{Bq}$.
 The real component of Eq. (\ref{Eq15}) can then be written as
 \begin{equation}
    A_{q} (t) = A_{q} (0) e^{-t/\tau_{q}} + \Bar{\zeta}_{Aq} (t)
    \label{Eq16}
\end{equation}
and a similar expression for $B_q$.

Therefore, it can be shown that
\begin{multline}
   \Big< |u_q (0)| |u_q(t)| \Big>  = \Big< |u_q (0)| \Big(A_q^2(t) +B_q^2(t)\Big)^{1/2} \Big> 
   \\ = \Big< |u_q (0)| \Big( |\Bar{\zeta}_{q}|^2(t)+2 \big(\Bar{\zeta}_{Aq}(t) A_q(0)+\Bar{\zeta}_{Bq} (t) B_q(0)\big)e^{-t/\tau_q} +|u_q(0)|^2e^{-2 t/\tau_q} \Big)^{1/2} \Big> \label{Eq17}
\end{multline}
  If we assume that $t>>t_q$, then we can perform the following expansion 
\begin{multline}  
   \Big< |u_q (0)| |u_q(t)| \Big> = \Big<   |u_q(0)|\Big>\Big<|\Bar{\zeta}_{q}(t)| \Big> \\+ \Bigg(  \Big<|u_q(0)| A_q(0)\Big>\Big< \frac{\Bar{\zeta}_{Aq}(t)}{|\Bar{\zeta}_{q}(t)|} \Big> + \Big<|u_q(0)| B_q(0)\Big>\Big< \frac{\Bar{\zeta}_{Bq}(t)}{|\Bar{\zeta}_{q}(t)|} \Big>  \Bigg) e^{-t/\tau_q} \\+ \frac{1}{2} \Bigg( \frac{\Big<|u_q(0)|^3\Big>}{\Big<|\Bar{\zeta}_{q}(t)|\Big>} - \Big< \frac{|u_q(0)|\big(\Bar{\zeta}_{Aq}(t) A_q(0)+\Bar{\zeta}_{Bq}(t) B_q(0)\big)^2}{|\Bar{\zeta}_{q}(t)|^3} \Big> \Bigg) e^{-2 t/\tau_q} + \mathcal{O}\big( e^{-3 t/\tau_q} \big).
   \label{Eq18}
\end{multline}

The second term in \eqref{Eq18} averages to zero due to the thermal noise factor. Therefore, to leading order, the CSD is given as

\begin{equation}
    \Big< |u_q (0)| |u_q(t)| \Big> - \Big< |u_q (0)| \Big>^2  = C_q e^{-2 t/\tau_q}  + \mathcal{O}\big( e^{-3 t/\tau_q} \big)
    \label{Eq19}
\end{equation}

where $C_q$ is a normalization constant.

\vspace{1mm}

Therefore, the slowest decaying mode of the CSD is $\mathcal{O}\big( e^{-2 t/\tau_q} \big)$. This contradicts H. Zhou et al. \cite{Losert:2011} who give it as $\mathcal{O}\big( e^{- t/\tau_q} \big)$. This factor of two is a consequence that each Fourier coefficient has both a real and imaginary component that are completely independent of each other. 

Finally, users are recommended to use time correlations over CSD. CSD requires the same amount of work and contains the same higher order error as the time correlation method. Yet, CSD has an additional layer of truncation error introduced in the expansion in \refeq{Eq18}.

\clearpage

\subsection*{Movie S1}

Real-time video of the GUV from Figure 1 in the main text (DOPC labeled with 0.2 mol $\%$ TR-DHPE) acquired with phase contrast microscopy. The vesicle radius is 29.6 $\mu$m.

\subsection*{Movie 2}

Video of the GUV from Figure 1 in the main text (DOPC labeled with 0.2 mol $\%$ TR-DHPE) acquired with confocal microscopy. The objective used is 40x/0.6 NA with the pinhole size 1 A.U. The polarization effect was corrected by using circular rotation plates to have even intensities across the equatorial vesicle plane. The vesicle radius is 29.6 $\mu$m.

\subsection*{Movie S3}

Real time video of the GUV from Figure 6 in the main text consisting of DOPC labeled with TR DHPE $(0.8\%)$ acquired with confocal microscopy. The objective used is 40x/0.6 NA at 13.2 fps with the pinhole size 1 A.U. The polarization effect was corrected by using circular rotation plates to have even intensities across the equatorial vesicle plane.

\subsection*{Raw data}

Raw data are available at \url{https://dx.doi.org/10.17617/3.4p}. This collection of raw data, consists of 4 folders each containing zipped files of data.\

In folder “Raw data - Fourier modes”, 2 different sets of experimental data are included: phase contrast (PC) and confocal (C) microscopy on the same vesicle and confocal microscopy data with different pinhole sizes. The folder contains excel sheets with fluctuation amplitude for every Fourier mode and the metadata with all the microscopy conditions. The vesicles have different sizes so that they practically cover a good span of focal depth ($\Delta$) from 0.03 to 0.15. The meta data is included in the first sheet of the excel file. The second sheet has the mode and mean squared amplitude (and error). The remaining sheets have the Fourier modes for every microscopy setting (and focal depth). Note that our Fourier signal was normalized by vesicle radius. For the definition of our Fourier transform, please refer to \textit{Gracia et al.}\cite{gracia.2010}. The contour detection is conducted as given in the main text and supplement. 

The rest of the folders as listed below contain vesicle images in tiff format (grouped in folders for the separate vesicle as suggested by the folder name) and an excel sheet with the meta data indicating the specific microscopy conditions (AU = Airy unit, NA = numerical aperture).

The folder “Raw images - different focal depth” contains confocal microscopy raw images at different focal depths without any image processing for vesicles of different sizes. 

The folder “Raw images - phase contrast vs confocal” contains phase contrast and confocal microscopy raw images for the same vesicle without any image processing for vesicles of different sizes. 

The folder “Raw images - polarization correction” contains polarized and polarization-corrected confocal microscopy raw images for the same vesicle without any image processing.

\pagebreak

\end{widetext}


\bibliography{rsc,refs2019,refsFluct,Refs} 

\end{document}